\documentclass[twocolumn]{aastex631}

\usepackage{multirow}

\begin{document}
\title{XMM-NEWTON CONCLUSIVELY IDENTIFIES AN ACTIVE GALACTIC NUCLEUS IN A GREEN PEA GALAXY\footnote{Based on observations obtained with XMM-Newton, an ESA science mission with instruments and contributions directly funded by ESA Member States and NASA}}

\author[0000-0001-9379-4716]{Peter~G.~Boorman}
\affiliation{Cahill Center for Astrophysics, California Institute of Technology, 1216 East California Boulevard, Pasadena, CA 91125, USA}
\affiliation{Astronomical Institute, Academy of Sciences, Bo\v{c}n\'{i} II 1401, CZ-14131 Prague, Czech Republic}

\author[0000-0003-2931-0742]{Ji\v{r}\'{i}~Svoboda}
\affiliation{Astronomical Institute, Academy of Sciences, Bo\v{c}n\'{i} II 1401, CZ-14131 Prague, Czech Republic}

\author[0000-0003-2686-9241]{Daniel~Stern}
\affiliation{Jet Propulsion Laboratory, California Institute of Technology, Pasadena, CA 91109, USA}

\author[0000-0003-2192-3296]{Bret~D.~Lehmer}
\affiliation{Department of Physics, University of Arkansas, 226 Physics Building, 825 West Dickson Street, Fayetteville, AR 72701, USA}
\affiliation{Arkansas Center for Space and Planetary Sciences, University of Arkansas, 332 N. Arkansas Avenue, Fayetteville, AR 72701, USA}

\author[0000-0002-9807-4520]{Abhijeet~Borkar}
\affiliation{Astronomical Institute, Academy of Sciences, Bo\v{c}n\'{i} II 1401, CZ-14131 Prague, Czech Republic}

\author[0000-0002-8147-2602]{Murray~Brightman}
\affiliation{Cahill Center for Astrophysics, California Institute of Technology, 1216 East California Boulevard, Pasadena, CA 91125, USA}

\author[0000-0001-5857-5622]{Hannah~P.~Earnshaw}
\affiliation{Cahill Center for Astrophysics, California Institute of Technology, 1216 East California Boulevard, Pasadena, CA 91125, USA}

\author[0000-0002-4226-8959]{Fiona~A.~Harrison}
\affiliation{Cahill Center for Astrophysics, California Institute of Technology, 1216 East California Boulevard, Pasadena, CA 91125, USA}

\author[0000-0002-1444-2016]{Konstantinos~Kouroumpatzakis}
\affiliation{Astronomical Institute, Academy of Sciences, Bo\v{c}n\'{i} II 1401, CZ-14131 Prague, Czech Republic}

\author[0009-0008-6899-4749]{Barbora~Adamcov\'{a}}
\affiliation{Astronomical Institute, Academy of Sciences, Bo\v{c}n\'{i} II 1401, CZ-14131 Prague, Czech Republic}

\author[0000-0002-9508-3667]{Roberto~J.~Assef}
\affiliation{Instituto de Estudios Astrof\'isicos, Facultad de Ingenier\'ia y Ciencias, Universidad Diego Portales, Av. Ej\'ercito Libertador 441, Santiago, Chile}

\author[0000-0002-2171-2926]{Matthias~Ehle}
\affiliation{European Space Agency, European Space Astronomy Centre (ESA/ESAC), Camino Bajo del Castillo s/n, 28692 Villanueva de la Ca\~{n}ada, Madrid, Spain}

\author[0000-0002-1984-2932]{Brian~Grefenstette}
\affiliation{Cahill Center for Astrophysics, California Institute of Technology, 1216 East California Boulevard, Pasadena, CA 91125, USA}

\author[0000-0003-3471-7459]{Romana~Grossov\'{a}}
\affiliation{Astronomical Institute, Academy of Sciences, Bo\v{c}n\'{i} II 1401, CZ-14131 Prague, Czech Republic}
\affiliation{Department of Theoretical Physics and Astrophysics, Faculty of Science, Masaryk University, Kotl\'{a}\v{r}sk\'{a} 2, Brno 61137, Czech Republic}

\author[0000-0003-0976-8932]{Maitrayee~Gupta}
\affiliation{Astronomical Institute, Academy of Sciences, Bo\v{c}n\'{i} II 1401, CZ-14131 Prague, Czech Republic}

\author[0000-0002-0273-218X]{Elias~Kammoun}
\affiliation{Cahill Center for Astrophysics, California Institute of Technology, 1216 East California Boulevard, Pasadena, CA 91125, USA}

\author[0000-0002-6808-2052]{Taiki~Kawamuro}
\affiliation{Department of Earth and Space Science, Osaka University, 1-1 Machikaneyama, Toyonaka 560-0043, Osaka, Japan}

\author[0000-0002-8472-3649]{Lea~Marcotulli}
\affiliation{Yale Center for Astronomy \& Astrophysics, 52 Hillhouse Avenue, New Haven, CT 06511, USA}
\affiliation{Department of Physics, Yale University, P.O. Box 208120, New Haven, CT 06520, USA}
\affiliation{Cahill Center for Astrophysics, California Institute of Technology, 1216 East California Boulevard, Pasadena, CA 91125, USA}

\author[0000-0001-7374-843X]{Romana~Miku\v{s}incov\'{a}}
\affiliation{INAF Istituto di Astrofisica e Planetologia Spaziali, Via del Fosso del Cavaliere 100, I-00133 Roma, Italy}

\author[0000-0002-8183-2970]{Matthew~J.~Middleton}
\affiliation{Department of Physics \& Astronomy, Faculty of Physical Sciences and Engineering, University of Southampton, Southampton, SO17 1BJ, UK}

\author[0000-0002-9633-9193]{Edward~Nathan}
\affiliation{Cahill Center for Astrophysics, California Institute of Technology, 1216 East California Boulevard, Pasadena, CA 91125, USA}

\author[0000-0003-1661-2338]{Joanna~M.~Piotrowska}
\affiliation{Cahill Center for Astrophysics, California Institute of Technology, 1216 East California Boulevard, Pasadena, CA 91125, USA}

\author[0000-0001-8426-5732]{Jean~J.~Somalwar}
\affiliation{Cahill Center for Astrophysics, California Institute of Technology, 1216 East California Boulevard, Pasadena, CA 91125, USA}

\author[0000-0003-3638-8943]{N\'{u}ria Torres-Alb\`{a}}
\altaffiliation{GECO Fellow}
\affiliation{Department of Astronomy, University of Virginia, P.O. Box 400325, Charlottesville, VA 22904, USA}

\author[0000-0001-5819-3552]{Dominic~J.~Walton}
\affiliation{Centre for Astrophysics Research, University of Hertfordshire, College Lane, Hatfield AL10 9AB, UK}

\author[0000-0002-6442-6030]{Daniel~R.~Weisz}
\affiliation{Department of Astronomy, University of California, Berkeley, CA 94720, USA}



\begin{abstract}
Green Pea galaxies are a class of compact, low-mass, low-metallicity star-forming galaxies in the relatively local universe. They are believed to be analogues of high-redshift galaxies that re-ionised the universe and, indeed, the \textit{James Webb Space Telescope} (\textit{JWST}) is now uncovering such populations at record redshifts. Intriguingly, \textit{JWST} finds evidence suggestive of active galactic nuclei (AGN) in many of these distant galaxies, including the elusive Little Red Dots, that broadly lack any detectable X-ray counterparts. Intuitively, one would expect to detect an AGN in their low-redshift analogues with X-rays, yet no study to date has conclusively identified an X-ray AGN within a Green Pea galaxy. Here we present the deepest X-ray campaign of a Green Pea galaxy performed to date, obtained with the goal of discerning the presence of a (potentially low-luminosity) AGN. The target -- SDSS\,J082247.66\,+224144.0 (J0822$+$2241 hereafter) -- was previously found to display a comparable X-ray spectral shape to more local AGN ($\Gamma$\,$\sim$\,2) and a high luminosity ($L_{2-10\,{\rm keV}}$\,$\sim$\,10$^{42}$\,erg\,s$^{-1}$). We show that over 6.2 years (rest-frame), the 2\,--\,10\,keV luminosity of J0822$+$2241 is constant, whereas the soft 0.5\,--\,2\,keV flux has decreased significantly by $\sim$\,60\%. We discuss possible scenarios to explain the X-ray properties of J0822$+$2241, finding transient low-column density obscuration surrounding an AGN to be the only plausible scenario. J0822$+$2241 thus provides further evidence that low-luminosity AGN activity could have contributed to the epoch of reionisation, and that local analogues are useful to derive a complete multi-wavelength picture of black hole growth in high redshift low luminosity AGN.
\end{abstract}




\section{Introduction}\label{sec:intro}

Early cosmic epochs (i.e., $z$\,$\gtrsim$\,6) were a crucial time in the history of the universe, corresponding to the period when energetic photons ionized and heated the intergalactic medium, leading to the end of the cosmic dark ages (e.g., \citealt{Bouwens15,Robertson15}). The source of such ionising photons is still a matter of debate. Neutral hydrogen readily absorbs and is ionised by ultraviolet photons. Thus the two most prominent sources of astrophysical ultraviolet flux -- accretion onto massive black holes and young populations of massive stars within compact low-mass galaxies -- are two prime candidates to have powered the epoch of reionisation (e.g., \citealt{Shapiro87,Loeb01,TorresAlba20}). X-ray photons could have also contributed to reionisation, with accretion onto massive black holes and X-ray binary populations being the two most likely astrophysical contenders \citep{Fragos13}. Though at ultraviolet and X-ray wavelengths many studies previously suggested that quasars should have been too rare at $z$\,$\gtrsim$\,6 to dominate cosmic reionisation (e.g., \citealt{Fontanot12,Fragos13,Haardt15}), numerous candidate active galactic nuclei (AGN) unveiled by the \textit{James Webb Space Telescope} (\textit{JWST}; \citealt{Gardner23}) have suggested accretion onto massive black holes in AGN with lower luminosities than powerful quasars may be a viable possibility after all (e.g., \citealt{Naidu22,Harikane23,Kocevski23,Yang23,Greene24,Uebler24,Madau24,Asthana24}).

\textit{JWST} has led to the discovery of many $z$\,$>$\,6 galaxies (including the so-called Little Red Dots; \citealt{Matthee23}) with broad permitted optical lines consistent with low-luminosity AGN. However, the lack of X-ray detections for the bulk of the population has led to ambiguity between extreme star formation processes and AGN (e.g., \citealt{Ananna24,Yue24,Maiolino25}). At lower redshifts, X-ray observations provide one of the most efficient methods for both selecting and characterising AGN (e.g., \citealt{Brandt15,Hickox18}). In particular, detailed X-ray spectral analyses have proven powerful in understanding the obscuration properties of AGN, including low-luminosity AGN (e.g., \citealt{Ricci15,Aird15,Buchner15,Annuar20,Civano24,Boorman24_hexp,Boorman24_ic750,Boorman25_nulands}). A unique perspective on the role of AGN versus star formation processes to cosmic reionisation is therefore attainable from nearby analogues of high redshift galaxies for which more detailed X-ray studies can be performed (e.g., \citealt{Svoboda19,Kawamuro19,Kouroumpatzakis24,Borkar24,Adamcova24,Singha24}).

Green Pea galaxies represent such a class of objects. First discovered by \citet{Cardamone09} from the Sloan Digital Sky Survey (SDSS; \citealt{York00}), Green Pea galaxies are now known to be compact (half-light radii\,$\lesssim$\,5\,kpc), low-mass ($M_{*}$\,$\lesssim$\,3\,$\times$\,10$^{9}$\,M$_{\odot}$), low metallicity (log\,[O\,/\,H]\,$+$\,12\,$\sim$\,8.1) star-forming galaxies with high star formation rates ($\gtrsim$\,10\,M$_{\odot}$\,yr$^{-1}$). Green Pea galaxies are also one of the closest (typically with $z$\,$\lesssim$\,0.3) galaxies known to exhibit significant Lyman continuum leakage to a level that is compatible with models of cosmic reionisation \citep{Izotov16}. Most recently, Green Pea galaxies displaying broad permitted lines have shown to be local analogs of \textit{JWST}-detected Little Red Dots due to their strikingly similar V-shaped rest-frame UV-to-optical spectra, compact morphologies and broad permitted lines akin to narrow line Seyfert~1 galaxies or regular broad line Seyfert~1 galaxies \citep{Lin25}.

However, detailed X-ray spectroscopic studies of Green Pea galaxies to infer the presence of an AGN have proven scarce to-date. The pioneering work of \citet{Svoboda19} revealed some unexpected X-ray properties for a sample of three Green Peas identified from the original \citet{Cardamone09} sample. Two of the three galaxies were found to be over-luminous by a factor of approximately five relative to empirical scaling relations that predict the level of X-ray luminosity expected as a function of metallicity and star formation (e.g., \citealt{Lehmer10,Brorby16}). This paper presents a detailed analysis of one Green Pea galaxy from the sample of \citet{Svoboda19}, SDSS\,J082247.66\,$+$224144.0 (denoted as J0822$+$2241 hereafter). At a redshift of $z$\,=\,0.216, detailed \textit{HST}/COS NUV observations confirmed a compact galaxy with a half light radius of 680\,pc \citep{Yang17}. The star formation rate, stellar mass and metallicity have additionally been estimated to be 37\,$\pm$\,4\,M$_{\odot}$\,yr$^{-1}$, $M_{*}$\,=\,3\,$\times$\,10$^{8}$\,M$_{\odot}$ and log(O/H)\,$+$\,12\,=\,8.1 respectively \citep{Kauffmann03,Brinchmann04,Cardamone09,Izotov11,Svoboda19}. However, the previous analyses of the SDSS optical spectrum considered a general lack of any detectable AGN component was present. Narrow line ratios plotted on the H$\alpha$/[N\textsc{ii}] versus H$\beta$/[O\textsc{iii}] Baldwin, Phillips and Terlevich (BPT) diagram \citep{Baldwin81} were consistent with theoretical expectations from star formation \citep{Svoboda19} and no common optical coronal emission lines indicative of massive black hole activity were significantly detected \citep{Reefe23}.

However, in X-rays J0822$+$2241 displayed a broad band X-ray continuum redolent of local AGN with an observed photon index 2.0\,$\pm$\,0.4 in the 0.3\,--\,10\,keV passband. The rest-frame 0.5\,--\,8\,keV X-ray luminosity of the source was additionally found to be substantial at $L_{0.5-8\,{\rm keV}}$\,$=$\,1.2$_{-0.3}^{+0.2}$\,$\times$\,10$^{42}$\,erg\,s$^{-1}$, compatible with the X-ray luminosities of local Seyfert AGN (e.g., \citealt{Ricci17_bassV,Annuar20}). \citet{Kawamuro19} showed that the near-to-mid infrared colour of J0822$+$2241 measured by the \textit{Wide-field Infrared Survey Explorer} (\textit{WISE}) was remarkably similar to the red colours found by powerful quasars and mid-infrared-dominated AGN in the more local universe (e.g., \citealt{Jarrett11,Stern12,Mateos12,Satyapal18,Assef18,Asmus20}). However as pointed out by \citet[see also \citealt{Sturm25}]{Hainline16}, the near-to-mid infrared colours expected from extreme star formation in compact galaxies with correspondingly high specific star formation rates can be arbitrarily red, in close similarity to the red colours expected from dominant AGN. Indeed, \citet{Kawamuro19} proved with extensive simulations that combined the same \textit{XMM-Newton} data analysed by \citet{Svoboda19} with non-detections at $>$\,10\,keV from snapshot observations with the \textit{Nuclear Spectroscopic Telescope ARray} (\textit{NuSTAR}; \citealt{Harrison13}) that either J0822$+$2241 is a Compton-thick Type~2 quasar observed in scattered light with a relatively unobscured spectral shape at $<$\,10\,keV, or that the near-to-mid infrared colours of the source cannot be reliably used as a bolometric indicator of AGN power. Additional insights were provided by \citet{Franeck22} who showed that hot gas could not explain the high X-ray luminosity of the source. \citet{Adamcova24} then calculated the expected X-ray emission from X-ray binaries in J0822$+$2241 by integrating a gas-phase metallicity-dependent X-ray luminosity function from \citet{Lehmer21}. By self-consistently accounting for star formation rate, metallicity and stochasticity effects, the authors showed that the observed 0.5\,--\,8\,keV luminosity of J0822$+$2241 could not have a contribution from X-ray Binaries greater than $\sim$\,20\%. J0822$+$2241 thus represents one of the strongest Green Pea X-ray AGN candidates identified to-date. However, a substantial 2\,--\,10\,keV luminosity and observed spectral index from a single relatively short exposure were insufficient to conclusively decipher its AGN nature. 

Here we present a detailed X-ray spectral and broadband investigation into J0822$+$2241 using 111\,ks of new data from two observations with the \textit{XMM-Newton} observatory (PI: P.~Boorman). Combined with the archival 28\,ks of \textit{XMM-Newton} data (PI: M.~Ehle), this study represents the deepest X-ray observation of a Green Pea galaxy performed to-date. In Section~\ref{sec:data} we present the X-ray observations of J0822$+$2241 together with a description of the X-ray, optical and broadband spectral methodology we use in this work. Section~\ref{sec:results} then presents the results of our multi-epoch X-ray spectral analysis of J0822$+$2241 followed by a discussion of its black hole and stellar masses in Section~\ref{sec:mstel} as well as viable scenarios for the physical origin of its X-ray properties in Section~\ref{sec:discussion}. Section~\ref{sec:jwstcomp} then compares the properties of J0822$+$2241 to a comparable sample of \textit{JWST}-detected AGN before a brief summary of our findings is given in Section~\ref{sec:summary}.

\section{Data and Method}\label{sec:data}

Details of all three \textit{XMM-Newton} \citep{Jansen01} observations used in this work are given in Table~\ref{tab:xmm_info}. All data were analysed with the Scientific Analysis System (\textsc{sas}; \citealt{Gabriel04}) \textsc{v.20.0.0}. The EPIC-pn \citep{Struder01} observation data files were processed using the \textsc{sas} command \textsc{epproc} to generate calibrated and concatenated events files. Intervals of background flaring activity were filtered using light curves generated in energy ranges recommended in the \textsc{sas} threads.\footnote{For more information, see \url{https://www.cosmos.esa.int/web/xmm-newton/sas-thread-epic-filterbackground}.} Corresponding images for the pn detector were generated using the command \textsc{evselect}, and source spectra were extracted from circular regions centered on the SDSS coordinates of J0822$+$2241 after accounting for any astrometric offsets by eye. Background regions of similar size to the source regions were defined following the \textit{XMM-Newton} Calibration Technical Note XMM-SOC-CAL-TN-0018 \citep{Smith22}, ensuring the distance from the readout node was similar to that of the source region. The EPIC-pn source and background spectra were then extracted with \textsc{evselect} with patterns less than four. Finally, response and ancillary response matrices were created with the \textsc{rmfgen} and \textsc{arfgen} tools. Each of the three EPIC-pn observations were performed with the Thin Filter in Full Frame mode. We do not use EPIC-MOS \citep{Turner01} data in this work since the improvement in sensitivity in combination with EPIC-pn was not substantial and contributed to an increase in computation time associated with simultaneously fitting all spectra together.

Both Epochs~2a and~2b in Table~\ref{tab:xmm_info} have consistent soft, hard and broad passband count rates. Thus we co-added the spectra using the \texttt{ftool} \texttt{addspec}\footnote{\url{https://heasarc.gsfc.nasa.gov/ftools/caldb/help/addspec.txt}}. All \textit{XMM-Newton} analysis presented hereafter thus refers to epoch~1 as the observation in 2013 and epoch~2 as the spectrum derived from co-adding the two observations in 2020.

\begin{table*}
\centering
\caption{\textit{XMM-Newton} data used in this work.\label{tab:xmm_info}}
\begin{tabular}{ccccccccccc}
\hline\hline\\[-0.33cm]
  Obs.\,ID &
  Label &
  Obs. start & $\mathcal{T}$ & $\mathcal{C}_{\rm soft}$ & $\mathcal{C}_{\rm hard}$ & $\mathcal{C}_{\rm broad}$ & $\mathcal{S}_{\rm soft}$ & $\mathcal{S}_{\rm hard}$ & $\mathcal{S}_{\rm broad}$ \\
(1) &
(2) &
(3)\,UT &
(4)\,ks &
(5)\,ct\,/\,ks &
(6)\,ct\,/\,ks &
(7)\,ct\,/\,ks &
(8) &
(9) &
(10) \\[0.2cm]
\hline\hline\\[-0.33cm]
0690470201 & \parbox[t]{1.3cm}{Epoch~1} & \parbox[t]{3cm}{2013-Apr-06,\,04:46} &          28.3 &        2.47\,$\pm$\,0.43 &        0.74\,$\pm$\,0.40 &         3.21\,$\pm$\,0.59 &                      5.9 &                      1.8 &                       5.5 \\
0865450301 & \parbox[t]{1.3cm}{Epoch~2a} & \parbox[t]{3cm}{2020-Oct-13,\,12:46} &          68.7 &        1.13\,$\pm$\,0.27 &        1.06\,$\pm$\,0.25 &         2.19\,$\pm$\,0.37 &                      4.2 &                      4.3 &                       6.0 \\
0865450401 & \parbox[t]{1.3cm}{Epoch~2b} & \parbox[t]{3cm}{2020-Nov-10,\,10:08} &          42.0 &        1.16\,$\pm$\,0.36 &        0.69\,$\pm$\,0.36 &         1.87\,$\pm$\,0.51 &                      3.2 &                      1.9 &                       3.7 \\
     \ldots & \parbox[t]{1.3cm}{Epoch~2\\(2a\,$+$\,2b)} &                                 \ldots &         110.7 &        1.14\,$\pm$\,0.17 &        0.92\,$\pm$\,0.16 &         2.07\,$\pm$\,0.23 &                      5.3 &                      4.4 &                       6.9 \\[0.5cm]
\hline\hline
\end{tabular}\\
{\raggedright \textbf{Notes.} (1)--observation ID; (2)--observation label used in this work; (3)--observation start date and time; (4)--net exposure time in ks; (5), (6) and (7)--net count rate in counts per ks for the soft (0.3\,--\,2\,keV), hard (2\,--\,10\,keV) and broad (0.3\,--\,10\,keV) bands, respectively; (8), (9) and (10)--signal-to-noise in the soft, hard and broad bands, respectively, computed with the \texttt{gv\_significance} library of \citet{Vianello18}.\footnote{\url{https://github.com/giacomov/gv_significance}}
}
\end{table*}

All X-ray spectral fitting presented in this paper was performed with \textsc{PyXspec} \citep{Gordon21,Arnaud96} using the modified C-statistic\footnote{\url{https://heasarc.gsfc.nasa.gov/xanadu/xspec/manual/XSappendixStatistics.html}} \citep{Cash79,Wachter79}. All parameter exploration was carried out with the Bayesian X-ray Analysis software package (\textsc{BXA} v4.0.5; \citealt{Buchner14,Buchner16}), using the nested sampling package \textsc{UltraNest} v4.0.5 \citep{Buchner21_ultranest}. All parameters were assigned uniform or log-uniform priors depending on their nature (i.e. whether it ranges over many orders of magnitude), unless stated otherwise. All spectral fits used a source redshift of $z$\,=\,0.216 and included Galactic absorption along the line-of-sight with column density $N_{\rm H}$\,=\,4.75\,$\times$\,10$^{20}$\,cm$^{-2}$ \citep{Willingale13} using the \texttt{TBabs} model and abundances from \citet{Wilms00}.

We rely on Quantile-Quantile plots for goodness-of-fit verification of the X-ray spectral fits in this work. Fundamentally Quantile-Quantile plots encompass the same information as more conventional residuals, in that the detected (source\,$+$\,background) counts are compared to the model-predicted counts as a means to understand if a given model can explain the data in an acceptable manner. However, for Quantile-Quantile plots, the detected and model-predicted counts are summed across the unbinned detector channels cumulatively (depicted as $\mathcal{Q}_{\rm data}$ and $\mathcal{Q}_{\rm model}$, respectively, throughout this work). For interpretative convenience, rather than relying on a plot presenting $\mathcal{Q}_{\rm data}$ vs. $\mathcal{Q}_{\rm model}$, we instead plot detected energy vs. $\mathcal{Q}_{\rm data}$\,--\,$\mathcal{Q}_{\rm model}$, more akin to conventional energy vs. data\,--\,model residuals (sometimes referred to as Quantile-Quantile difference plots; \citealt{Buchner23,Boorman24_ic750}). There are important distinctions to consider in comparison to standard data\,--\,model residuals, though. For example, a peak or trough present in energy vs. $\mathcal{Q}_{\rm data}$\,--\,$\mathcal{Q}_{\rm model}$ informs us that the largest data excess relative to the model or largest model excess relative to the data occurs below the energy of that peak or trough, respectively. There are also a number of advantages to performing model verification in terms of quantiles rather than standard residuals. First, Quantile-Quantile plots sum counts on the intrinsic detector energy resolution and do not require binning. Therefore in the event of low signal-to-noise data, valuable inference can still be acquired without requiring any loss of information. Second, by simulating a given model fit with the instrumental setup of the detector (i.e. the same background, response and exposure time), one can plot the predicted range in $\mathcal{Q}_{\rm data}$\,--\,$\mathcal{Q}_{\rm model}$ expected from the imperfect nature of the detector in the event that the model were correct. The corresponding posterior predictive range can therefore be used to quantify when fluctuations in $\mathcal{Q}_{\rm data}$\,--\,$\mathcal{Q}_{\rm model}$ are statistically significant. All Quantile-Quantile difference plots in this work provide the 90\% posterior predictive range as dark grey shaded regions. We consider any deviation in $\mathcal{Q}_{\rm data}$\,--\,$\mathcal{Q}_{\rm model}$ that is significantly outside a given posterior predictive range to be significant to $\geq$\,90\% confidence.

To complement our X-ray spectral fitting we re-analyse the archival SDSS spectrum of J0822$+$2241 (SDSS \texttt{SpecObjID} 2168502517682432000) to primarily search for a broad component to the H$\alpha$ line. We find a signal-to-noise ratio of $\gtrsim$\,70 and $\sim$\,5\,--\,7 in H$\alpha$ and the continuum over the 6400\,--\,6700\,${\rm \AA}$ passband, respectively. To analyse the spectrum, we use BXA\,v2.10 with the nested sampling package \texttt{PyMultiNest} \citep{Feroz09,Buchner14} within \texttt{PyXspec}. To load the spectrum into \texttt{Xspec}, we use the HEASoft tool \texttt{ftflx2xsp} to convert the SDSS spectrum into an \texttt{Xspec}-readable format. All line luminosities reported from the optical spectral fitting have been corrected for Milky Way extinction using the nebular colour excess estimation from \citet{Schlafy11}\footnote{Acquired via the NASA/IPAC Galactic Dust Reddening and Extinction tool; \url{https://irsa.ipac.caltech.edu/applications/DUST/}} and extinction law of \citet{Fitzpatrick99} via the \texttt{extinction} Python package\footnote{\url{https://extinction.readthedocs.io/en/latest/}}. The total estimated reddening arising along the line-of-sight to J0822$+$2241 from the Milky Way is $E(B-V)$\,=\,0.039\,mag, amounting to a multiplicative flux correction factor of 1.06 and 1.11 at the observed frame wavelengths of H$\alpha$ and H$\beta$, respectively.

To obtain independent estimates of the galaxy and AGN properties, we performed broadband spectral energy distribution (SED) fitting across the X-ray to infrared wavelength regime using the {\tt Lightning} package \citep{Doore23,Lehmer24}.\footnote{https://github.com/ebmonson/lightningpy.} Following the procedures outlined in Section~3 of \citet{Lehmer24}, we culled imaging data available in the archives from various facilities, including {\it GALEX}, {\it Swift}/UVOT, SDSS, PanSTARRS, and {\it WISE},\footnote{{\it GALEX} and {\it Swift}/UVOT data were obtained from the Mikulski Archive for Space Telescopes (MAST; \url{https://mast.stsci.edu/}), SDSS from DR18 via SkyServer (\url{https://skyserver.sdss.org/}), PanSTARRS from the PS1 Image Access portal (\url{https://ps1images.stsci.edu/cgi-bin/ps1cutouts}), and {\it WISE} from the NASA/IPAC Infrared Science Archive (IRSA; \url{https://irsa.ipac.caltech.edu/})} and convolved all data to a common 20\,arcsec Full Width Half Maximum (FWHM) Point Spread Function. Photometry was subsequently extracted from all bands using a circular aperture with a 33\,arcsec radius. We expanded our SED to include the {\it XMM-Newton} count spectrum for epoch~2 in four additional X-ray bandpasses. In total, our broadband SED includes 19 photometric measurements spanning the X-ray to mid-infrared. When fitting the SED with {\tt Lightning}, we followed the methods detailed in \citet{Monson23}, assuming that both stellar populations and AGN are able to contribute. In this framework, the AGN is modelled using {\tt qsosed} \citep{Kubota2018} to account for the intrinsic accretion disk spectral shape, and {\tt SKIRTOR} \citep{Stalevaski2012,Stalevski2016} to model inclination-dependent dust obscuration and emission from a dusty torus. Included among the full set of parameters (see Table~4 of \citealt{Monson23}) are the star-formation history in five discrete age bins, which enables the derivation of star formation rate, galaxy stellar mass and the central black hole mass of the AGN.

All spectral parameters throughout the paper are quoted as the maximum a posteriori value together with the 90\% highest density interval integrated from the marginalised posterior mode, unless stated otherwise. We additionally use $u$ to describe unconstrained parameter bounds. Luminosities are calculated assuming the cosmological parameters $H_{0}$\,=\,70\,km\,s$^{-1}$\,Mpc$^{-1}$, $\Omega_{\rm m}$\,=\,0.3, $\Omega_{\Lambda}$\,=\,0.7, which correspond to a luminosity distance to J0822$+$2241 at $z$\,=\,0.216 of 1.07\,Gpc.

\section{Results}\label{sec:results}

\begin{table*}
\centering
\caption{X-ray spectral parameters constrained in this work.\label{tab:par_info}}
\begin{tabular}{ccccc}
\hline\hline\\[-0.33cm]
Parameter &
Passband &
Epoch~1 &
Epoch~2 &
Units \\
(1) &
(2)\,keV &
(3) &
(4) &
(5) \\[0.2cm]
\hline\hline\\[-0.33cm]
\multicolumn{5}{c}{Phenomenological model parameters (c.f.~Section~\ref{subsec:xrayres})}\\
\multicolumn{5}{c}{\texttt{model}\,=\,\texttt{TBabs}\,$\times$\,\texttt{zpowerlw}}\\[0.2cm]
\hline\\[-0.33cm]
$\Gamma$ &
\ldots &
$1.7_{-0.6}^{+0.5}$ &
$0.8_{-0.3}^{+0.2}$ &
\ldots \\
log\,$F_{\rm obs}$$^{a}$ &
0.5\,--\,2 &
$-14.4$\,$\pm$\,0.1 &
$-14.7$\,$\pm$\,0.1 &
erg\,s$^{-1}$\,cm$^{-2}$ \\
log\,$F_{\rm obs}$$^{a}$ &
2\,--\,10 &
$-14.0_{-0.4}^{+0.3}$ &
$-13.8$\,$\pm$\,0.1  &
erg\,s$^{-1}$\,cm$^{-2}$ \\
log\,$L_{\rm obs}$$^{b}$ &
0.5\,--\,2 &
$41.7$\,$\pm$\,0.1 &
$41.4$\,$\pm$\,0.1 &
erg\,s$^{-1}$ \\
log\,$L_{\rm obs}$$^{b}$ &
2\,--\,10 &
$42.0$\,$\pm$\,0.3 &
$42.2$\,$\pm$\,0.1 &
erg\,s$^{-1}$ \\[0.2cm]
\hline\\[-0.33cm]
\multicolumn{5}{c}{Obscured AGN physical model parameters (c.f.~Section~\ref{subsec:obscuredagn})}\\
\multicolumn{5}{c}{\texttt{model}\,=\,\texttt{TBabs}\,$\times$\,\texttt{zTBabs}\,$\times$\,\texttt{cabs}\,$\times$\,\texttt{zpowerlw}}\\[0.2cm]
\hline\\[-0.33cm]
log\,$N_{\rm H}$ &
\ldots &
$20.7_{-u}^{+0.7}$ &
$21.8$\,$\pm$\,0.3 &
cm$^{-2}$ \\
log($N_{\rm H,\,2}$\,/\,$N_{\rm H,\,1}$)$^{c}$ &
\ldots &
\multicolumn{2}{c}{1.08$_{-0.72}^{+0.66}$} &
\ldots \\
$\Gamma$ &
\ldots &
\multicolumn{2}{c}{$1.7_{-0.2}^{+0.1}$} &
\dots\\
log($L_{\rm 1,\,obs}$\,/\,$L_{\rm 2,\,obs}$)$^{d}$ &
0.5\,--\,2 &
\multicolumn{2}{c}{0.22$_{-0.12}^{+0.13}$} &
\ldots \\
log($L_{\rm 1,\,obs}$\,/\,$L_{\rm 2,\,obs}$)$^{d}$ &
2\,--\,10 &
\multicolumn{2}{c}{0.01\,$\pm$\,0.01} &
\ldots \\
log\,$L_{\rm int}$$^{e}$ &
0.5\,--\,2 &
\multicolumn{2}{c}{$41.8$\,$\pm$\,0.1} &
erg\,s$^{-1}$ \\
log\,$L_{\rm int}$$^{e}$ &
2\,--\,10 &
\multicolumn{2}{c}{$42.1$\,$\pm$\,0.1} &
erg\,s$^{-1}$ \\[0.2cm]
\hline\hline
\end{tabular}\\
{\raggedright \textbf{Notes.} (1)--parameter of interest; (2)--passband that a corresponding parameter was measured over; (3), (4)--observed parameter value measured for the epoch~1 and~2 X-ray spectra, respectively; (5)--units of the parameter of interest. $^{a}$absorption-uncorrected observed-frame flux; $^{b}$absorption-uncorrected rest-frame luminosity; $^{c}$logarithmic column density ratio between epoch~2 and~1; $^{d}$logarithmic absorption-uncorrected rest-frame luminosity ratio between epoch~1 and~2; $^{e}$absorption corrected rest-frame luminosity.
}
\end{table*}

\subsection{X-ray Spectral Analysis}\label{subsec:xrayres}

\begin{figure*}
\centering
\includegraphics[width=0.99\textwidth]{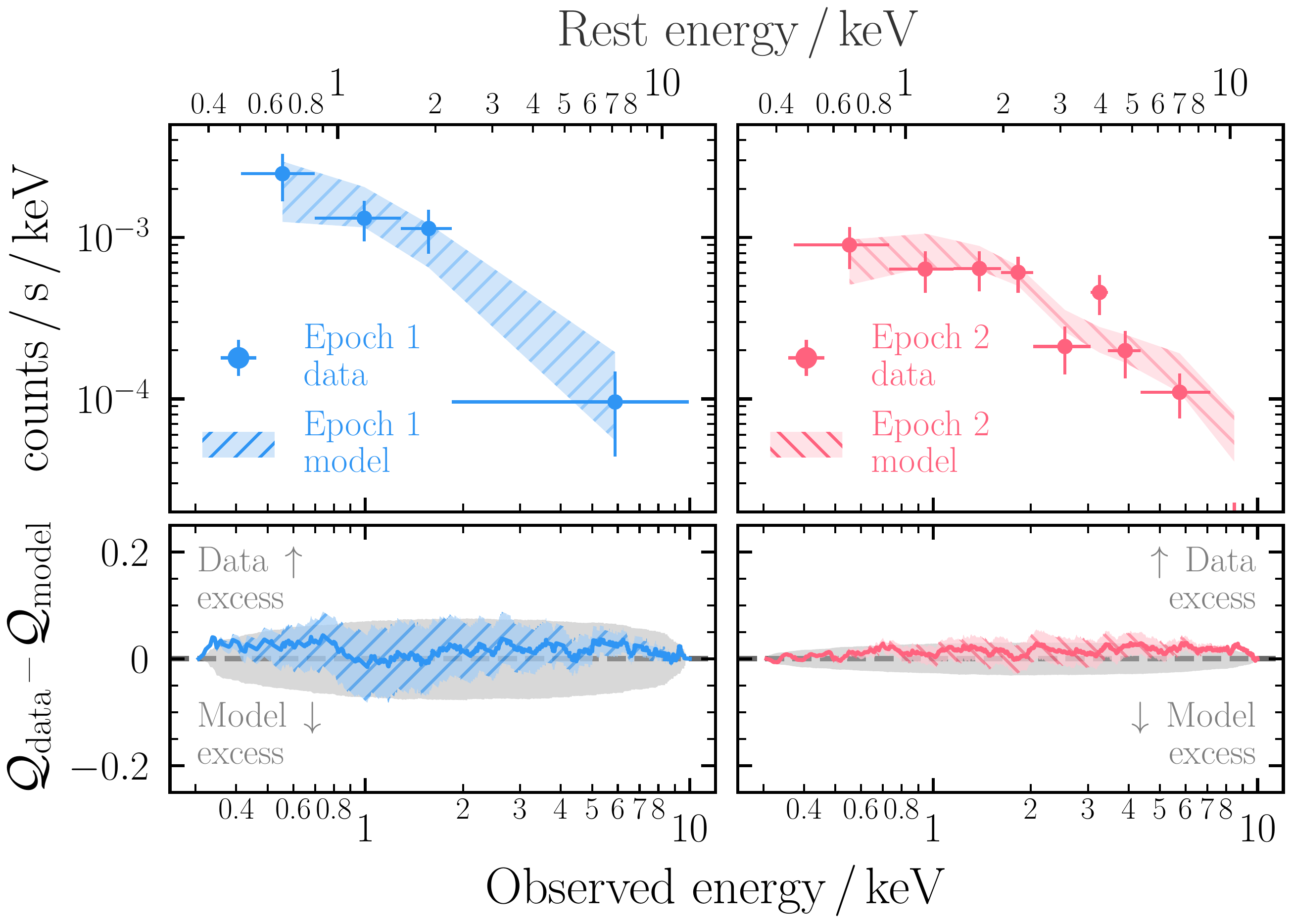}
\caption{Independent spectral fits to epoch~1 (left) and epoch~2 (right) \textit{XMM-Newton} data of J0822$+$2241. The model used for both epochs is a redshifted powerlaw with fixed Galactic absorption, and the hatched shaded regions show the 90\% model posterior uncertainty in all panels. The observed spectral shape is harder in epoch~2 than epoch~1, driven predominantly by a drop in flux at energies $\lesssim$\,2\,keV. The lower panels show the Quantile-Quantile difference plots in which $\mathcal{Q}_{\rm data}$\,--\,$\mathcal{Q}_{\rm model}$ are plotted against energy (see Section~\ref{sec:data} for more information). The background grey shaded regions represent the 90\% posterior predictive range derived by simulating a random selection of the posterior model rows many times with the same instrumental setup as the real data (i.e. the same background, response and exposure time). Since the hatched shaded regions for both epochs agree with the grey shaded regions, we confirm that each model can explain the observed data in each epoch satisfactorily.}
\label{fig:gp2_zpow}
\end{figure*}

First, we focus on a phenomenological X-ray spectral parametrisation of J0822$+$2241 in both \textit{XMM-Newton} epochs separately. The upper portion of Table~\ref{tab:par_info} and Figure~\ref{fig:gp2_zpow} shows the spectral fits with a redshifted powerlaw to epochs~1 and~2. For epoch~1, we find an observed photon index of 1.7\,$^{+0.5}_{-0.6}$, consistent with the value of 2.0\,$\pm$\,0.4 derived by \citet{Svoboda19}. Additionally in agreement with \citeauthor{Svoboda19}, we find a substantial rest-frame absorption-uncorrected 2\,--\,10\,keV luminosity of log\,$L_{2-10\,{\rm keV}}$\,/\,erg\,s$^{-1}$\,=\,$42.0$\,$\pm$\,0.3 for J0822$+$2241 in epoch~1. The second epoch of data, taken $\sim$\,6.2\,years later in the rest-frame of J0822$+$2241, provides a means to search for X-ray variability in the target. We find that the hard 2\,--\,10\,keV X-ray rest-frame absorption-uncorrected luminosity of J0822$+$2241 is fully consistent with being constant between both epochs. However, the 0.5\,--\,2\,keV luminosities and associated uncertainties suggest a decrease of $\sim$\,0.3\,dex between epoch~1 and~2. The corresponding decrease in soft flux also results in a very hard observed photon index in epoch~2 of $\Gamma$\,=\,0.8$^{+0.2}_{-0.3}$, significantly outside the typical distribution of intrinsic (i.e. absorption-corrected) photon indices found for local AGN samples that tends to peak at $\sim$\,1.6\,--\,1.9 (e.g., \citealt{Nandra94,Ricci17_bassV}).

\begin{figure*}
\centering
\includegraphics[width=0.99\textwidth]{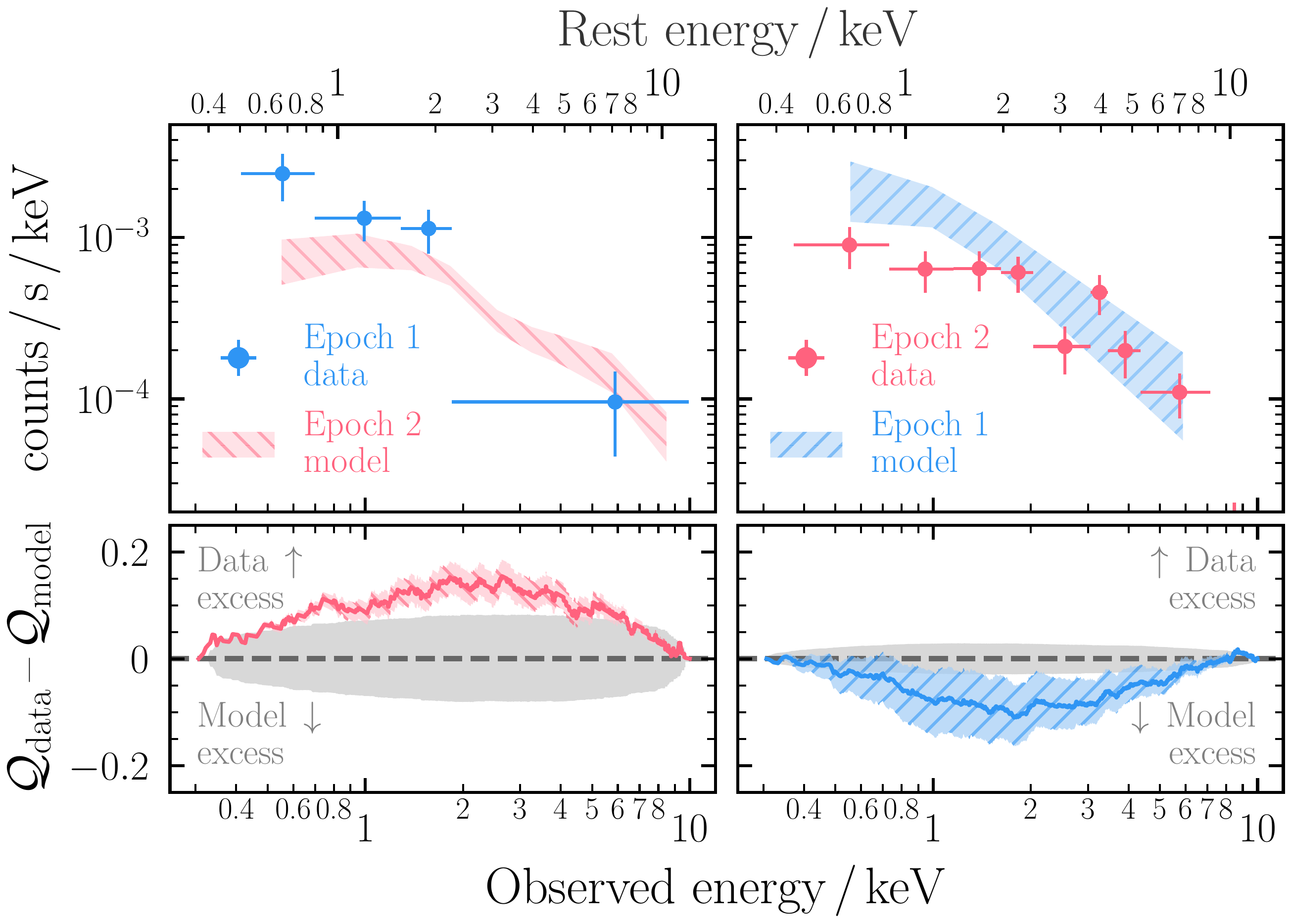}
\caption{The same as Figure~\ref{fig:gp2_zpow}, apart from the posterior model found for each epoch has been swapped to check if the epoch~1 model can explain the data from epoch~2 and vice-versa. The lower panels that show the Quantile-Quantile difference plots are discrepant to the posterior-predictive range shown with grey shading. The hatched shaded bands give a peak and trough at $\sim$\,2\,keV in the left and right lower panels, respectively. Since the information conveyed is cumulative, the lower left panel indicates there is a significant excess in detected counts for epoch~1 relative to the model posterior prediction from epoch~2 (see Section~\ref{sec:data} for more information). The opposite is true for the lower right panel, in which the detected counts in epoch~2 are significantly suppressed relative to the model posterior from epoch~1. Given the excesses relative to the grey shaded regions in either panel, we conclude that the deviations $\lesssim$\,2\,keV are significant to $>$\,90\% confidence.}
\label{fig:gp2_zpow_switch}
\end{figure*}

To understand if the apparent spectral change between either epoch is significant, we check if the posterior model derived for epoch~1 can explain the data from epoch~2 satisfactorily and vice-versa. Figure~\ref{fig:gp2_zpow_switch} presents the same folded X-ray spectral data as in Figure~\ref{fig:gp2_zpow}, apart from the posterior models from either epoch have been switched. From the upper panels alone, it is clear that either model cannot explain the observed data $\lesssim$\,2\,keV from the opposite epoch. In the lower panels, we show the Quantile-Quantile difference curves as a means to quantify the significance of the X-ray spectral shape change $\lesssim$\,2\,keV. Since the grey shaded posterior predictive regions in either lower panel rely on the same data as in Figure~\ref{fig:gp2_zpow}, the resulting grey shaded regions are very similar to Figure~\ref{fig:gp2_zpow} as well. For the lower left panel of Figure~\ref{fig:gp2_zpow_switch}, the $\mathcal{Q}_{\rm data}$\,--\,$\mathcal{Q}_{\rm model}$ curve presents a peak at $\sim$\,2\,keV significantly offset from the 90\% shaded posterior predictive region. Since the curve is cumulative, we can infer that the largest epoch~1 data excess relative to the epoch~2 posterior model is $\lesssim$\,2\,keV, and significant to $>$\,90\% confidence. The opposite is true in the lower right panel of Figure~\ref{fig:gp2_zpow_switch}, in which the epoch~2 data is suppressed relative to the epoch~1 model posterior to $>$\,90\% confidence. We note that in generating the $\mathcal{Q}_{\rm data}$\,--\,$\mathcal{Q}_{\rm model}$ curves and posterior predictive ranges, we deliberately simulate multiple posterior rows multiple times rather than a single best-fit in order to encompass the full possible posterior spectral range allowed by either fit. To our knowledge, this is the first statistically significant detection of X-ray spectral variability from a Green Pea galaxy. However, these X-ray spectral tests are purely phenomenological and do not necessarily provide a causal link between each epoch. We defer the reader to Section~\ref{subsec:obscuredagn} for a physically-plausible cause of this flux change.

\subsection{Optical Spectral Analysis}\label{subsec:optres}

Complimentary to our X-ray analysis, we additionally search for signatures of an AGN in J0822$+$2241 via spectral fitting of the archival SDSS spectrum (c.f.~Figure~\ref{fig:gp2_halpha}). Our fitting procedure was primarily constructed to follow that of \citet{Reines13}, in which a fitting procedure was devised to search for faint though statistically-significant evidence of broad components to H$\alpha$. To account for the underlying continuum produced from different stellar populations, we first used the penalized PiXel Fitting software (pPXF; \citealt{Cappellari04}) which includes emission from host galaxy starlight including Balmer absorption lines. However, there were no strongly detectable absorption features in the observed SDSS passband for pPXF to constrain stellar kinematics and/or the stellar velocity dispersion significantly. Due to the expected intense star-forming activity of J0822$+$2241, its spectrum is expected to be dominated by its ionised interstellar medium with a negligible contribution from older stellar populations. Thus, we do not perform starlight subtraction since no significant absorption features are expected. After experimenting with a number of alternative models for the underlying continuum, we settled for a simple redshifted power law model within \texttt{PyXspec} to constrain the pseudo continuum. Since the pseudo-continuum model chosen is not physical, and (as noted in Section~\ref{sec:data}) the signal-to-noise ratio of the continuum is far lower than the emission lines, we do not attempt to subtract the continuum to generate an emission line-only spectrum. Instead, we leave our pseudo-continuum model free to vary during all emission line fits, so that any emission line parameter uncertainties naturally incorporate the uncertainty associated with the pseudo-continuum itself. We note that all reported Balmer emission line fluxes could thus be under-estimated in the event that substantial Balmer absorption is present, which is not accounted for with our pseudo-continuum model. For all emission line fits, we used wide uniform and log-uniform priors for the pseudo-continuum power law photon index and normalisation with the \texttt{zpowerlw} model in \textsc{PyXspec}.

\subsubsection{Constructing a Narrow Line Template}\label{subsubsec:sii}
On visual inspection, the H$\alpha$\,$\lambda$6563 emission line appears to be significantly blended with the [N\textsc{ii}]\,$\lambda\lambda$6548,\,6583 doublet in the SDSS spectrum of J0822$+$2241. Thus to de-blend the [N\textsc{ii}] doublet from H$\alpha$, we follow the technique of \citet[see also  \citealt{Filippenko88,Filippenko89,Ho97,Greene04,Dong12,Liu25}]{Reines13} by fitting the [S\textsc{ii}]\,$\lambda\lambda$6716,\,6731 doublet to produce a narrow component template for each of the narrow [N\textsc{ii}] lines, as well as the narrow component to the H$\alpha$. We trial two narrow line models to fit [S\textsc{ii}]. The first contains a single Gaussian model per [S\textsc{ii}] line (allowed to vary in width between 50\,--\,300\,km\,s$^{-1}$), and the second contains two narrow Gaussian models (with the additional second Gaussians allowed to vary in width between 50\,--\,1000\,km\,s$^{-1}$). It is important to note that some previous Gaussian decompositions of the [S\textsc{ii}] complex within AGN SDSS spectra have required varying intensity ratios between each [S\textsc{ii}] line (see e.g., \citealt{Ho97} for examples), though typically in a minority of cases. However, we did not find such additional complexity was required whilst fitting the [S\textsc{ii}] doublet of J0822$+$2241, and thus the intensities of each [S\textsc{ii}] line were tied together in all corresponding parametrisations.

We fit the [S\textsc{ii}] doublet over the rest-frame wavelength range 6620\,--\,6850\,${\rm \AA}$\footnote{On visual inspection, we identified a prominent emission line coincident with He\textsc{i}\,$\lambda$6678. Thus for all fits encompassing this emission line, we excised the rest-frame 6678\,$\pm$\,14\,${\rm \AA}$ window from the spectrum.} with a total (including the two pseudo-continuum model parameters) of five and seven free parameters in the 1--Gaussian and 2--Gaussian models, respectively. For the 1--Gaussian model we varied a single normalisation and line width applied to each [S\textsc{ii}] line with log-uniform priors as well as the line centroid of [S\textsc{ii}]\,$\lambda$6716 with a uniform prior whilst enforcing that the relative separation of [S\textsc{ii}]\,$\lambda$6716 and [S\textsc{ii}]\,$\lambda$6731 was fixed to the laboratory value. The same priors were applied to the additional Gaussian lines used in the 2--Gaussian model, though with a wider allowable range in line width as specified above. The line centroid shift of both additional Gaussian lines was also tied to the value derived with the first Gaussian line model for [S\textsc{ii}]\,$\lambda$6716.

\begin{figure}
\centering
\includegraphics[width=0.99\columnwidth]{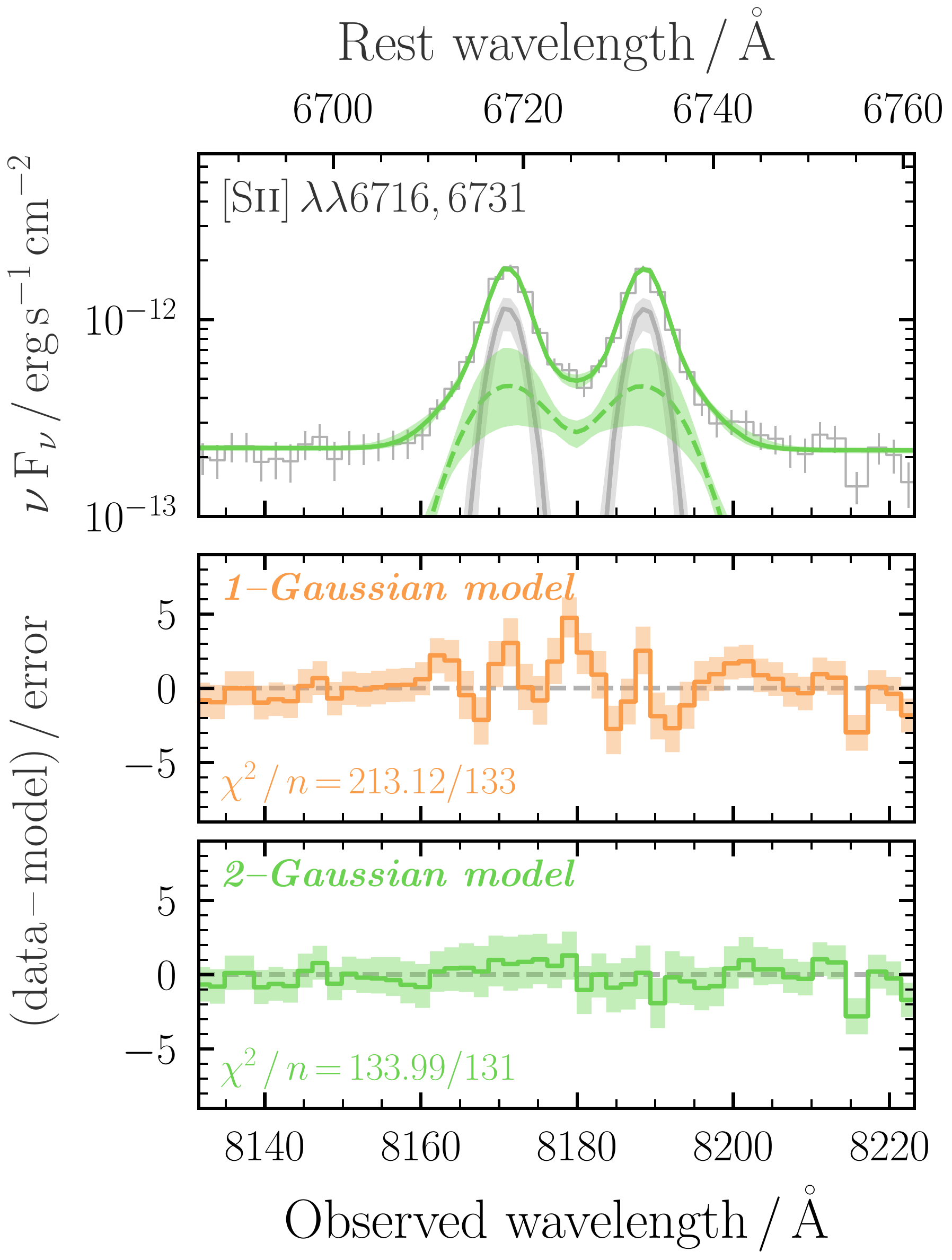}
\caption{(Top) [S\textsc{ii}]\,$\lambda\lambda$6716,\,6731 doublet in the SDSS spectrum of J0822$+$2241, fit with the 2--Gaussian spectral model described in Section~\ref{subsubsec:sii}. The green solid line shows the total model, whereas the grey solid lines and green dashed line show the first and second Gaussian model components to each [S\textsc{ii}] line, respectively. (Center) Residuals from a separate fit with the 1--Gaussian model in which a single Gaussian component was used to explain each [S\textsc{ii}] line. (Bottom) Residuals arising from the model with the 2--Gaussian model shown in the top panel. Given the improvement in reduced chi-squared, we use the 2--Gaussian model as a narrow emission line template for the [N\textsc{ii}]\,$\lambda\lambda$6548,\,6583 doublet and narrow core of the H$\alpha$\,$\lambda$6563 emission line.}
\label{fig:sii_fit}
\end{figure}

Figure~\ref{fig:sii_fit} presents the results from our spectral fits to the [S\textsc{ii}]\,$\lambda\lambda$6716,\,6731 complex of J0822$+$2241 with our 1--Gaussian and 2--Gaussian models. We find that a single Gaussian line is incapable of explaining the relatively broad base of either [S\textsc{ii}] emission line (see center panel of Figure~\ref{fig:sii_fit}). The corresponding chi-squared values for each spectral fit also favoured the inclusion of two Gaussian lines to explain the [S\textsc{ii}] doublet. We find an improvement in chi-squared from 213.12 with 133 degrees of freedom ($\chi^{2}_{n}$\,=\,1.60) to 133.99 with 131 degrees of freedom ($\chi^{2}_{n}$\,=\,1.02) when using the 1--Gaussian and 2--Gaussian models to explain the [S\textsc{ii}], respectively. The resulting 2--Gaussian model fit, highlighting the additional broader Gaussian component per [S\textsc{ii}] line, is shown in the upper panel of Figure~\ref{fig:sii_fit}. Given the general lack of significant residuals in the lower panel of Figure~\ref{fig:sii_fit}, we proceed with the 2--Gaussian model as a narrow line template for the [N\textsc{ii}]\,$\lambda\lambda$6548,\,6583 doublet as well as the narrow core of the H$\alpha$\,$\lambda$6563 line.

\subsubsection{Characterising the H$\alpha$ Complex}\label{subsubsec:halpha}
As shown in the upper panel of Figure~\ref{fig:sii_fit}, both components within the 2--Gaussian line template model display noticeable uncertainty within the posteriors of their model components. To self-consistently propagate all the information encompassed by these uncertainties into our final H$\alpha$\,$\lambda$6563 emission line constraints, we simultaneously fit the narrow line template with the [S\textsc{ii}]\,$\lambda\lambda$6716,\,6731 doublet in combination with the [N\textsc{ii}]\,$\lambda\lambda$6548,\,6583 doublet and H$\alpha$\,$\lambda$6563 emission line over the rest-frame 6400\,--\,7000\,${\rm \AA}$ passband. We incorporate two models in total: (i) a baseline model in which the [N\textsc{ii}] and [S\textsc{ii}] doublets and H$\alpha$\,$\lambda$6563 emission line are explained purely with our 2--Gaussian narrow line template derived in Section~\ref{subsubsec:sii}, and (ii) the same baseline model with an additional broad Gaussian line component to the H$\alpha$ emission line included. The relative scaling between the two Gaussian components of each narrow line was tied to that of the [S\textsc{ii}] doublet. To account for any asymmetries in the narrow H$\alpha$ emission line, we additionally allowed the line centroids of both components to vary uniformly by $\pm$\,10\,${\rm \AA}$. For the additional broad Gaussian line in model (ii), we allowed the normalisation, width and line centroid to vary as free parameters.

\begin{figure}
\centering
\includegraphics[width=0.99\columnwidth]{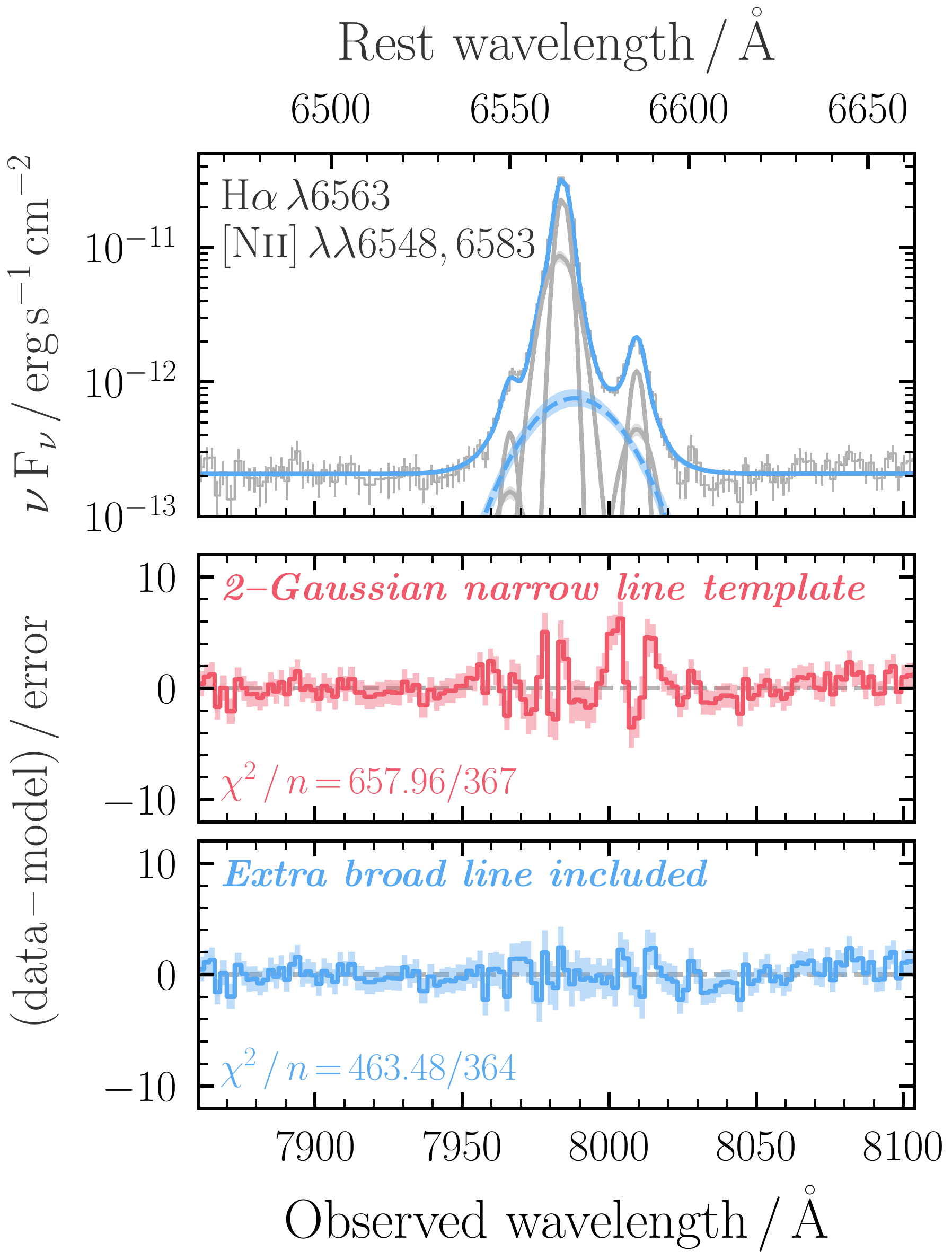}
\caption{(Top) SDSS H$\alpha$ and [N\textsc{ii}] complex of J0822$+$2241, with the spectral fit described in Section~\ref{subsubsec:halpha}. The solid blue line shows the total model, and the dashed blue line with shaded region shows the posterior constraint on the broad component to H$\alpha$. The grey lines and associated shading show the posterior constraints on the narrow components to the [N\textsc{ii}] doublet and H$\alpha$ line using the narrow line template derived in Section~\ref{subsubsec:sii}. (Center) Residuals from a similar fit in which no broad component to H$\alpha$ was included. (Bottom) Residuals arising from the model with a broad H$\alpha$ line included, relevant to the spectral fit shown in the top panel.}
\label{fig:gp2_halpha}
\end{figure}

Figure~\ref{fig:gp2_halpha} presents the results from our spectral fits with model (i) and (ii), focused on a zoom-in of the [N\textsc{ii}] and H$\alpha$ emission line complex. We found that the inflection points between H$\alpha$ and each [N\textsc{ii}] line were not well reproduced using just the narrow line template, resulting in strong residuals either side of H$\alpha$ (c.f. center panel of Figure~\ref{fig:gp2_halpha}). Using the narrow line template resulted in a chi-squared of 657.96 with 367 degrees of freedom ($\chi^{2}_{n}$\,=\,1.79). The inclusion of an additional broad emission line component to H$\alpha$ in model (ii) provided a significant improvement to the spectral fit with a chi-squared of 463.48 with 364 degrees of freedom ($\chi^{2}_{n}$\,=\,1.27). The upper and lower panels of Figure~\ref{fig:gp2_halpha} present the spectral fit with model (ii) and its corresponding residuals, respectively. The presence of the broad component significantly reduced the residuals either side of the narrow component to H$\alpha$, as expected. We also find that the observed flux posterior of the component is mono-modal and well constrained with log\,F$_{{\rm H}\alpha,\,{\rm broad,\,obs}}$\,/\,erg\,s$^{-1}$\,cm$^{-2}$\,=\,-14.41$^{+0.06}_{-0.05}$, implying the component is significantly required within the parameterisation of model (ii). The corresponding observed luminosity of the broad component to H$\alpha$ is log\,$L_{{\rm H}\alpha,\,{\rm broad},\,{\rm obs}}$\,/\,erg\,s$^{-1}$\,=\,41.72$^{+0.06}_{-0.05}$ with a FWHM$_{{\rm H}\alpha,\,{\rm broad}}$\,=\,1360$^{+70}_{-100}$\,km\,s$^{-1}$, after correcting for Milky Way extinction (but before accounting for extinction intrinsic to J0822$+$2241 -- see Section~\ref{subsubsec:hbeta}).

Broad H$\alpha$ with FWHM$_{{\rm H}\alpha,\,{\rm broad}}$\,$\lesssim$\,2000\,km\,s$^{-1}$ has been shown previously to trace the broad line region surrounding low mass AGN (e.g., \citealt{Reines13,Reines15}). However broad components to Balmer lines are also known to be prevalent in low mass and/or low metallicity systems similar to J0822$+$2241, often attributed to the interaction between massive stars and their interstellar medium and/or supernova activity (e.g., \citealt{Izotov07}). Thus care should always be taken in interpreting broad Balmer line components as purely AGN-driven (see also \citealt{Maiolino25} for a recent overview). In the cases in which broad Balmer line emission is powered by a dominant contribution from massive stars, the equivalent width of the broad component is typically $\lesssim$\,20\,${\rm \AA}$ \citep{Izotov07,Martins20}. For our spectral fit to J0822$+$2241, the equivalent width of the broad component to H$\alpha$ is EW$_{{\rm H}\alpha,\,{\rm broad}}$\,=\,90$^{+12}_{-10}$\,${\rm \AA}$, suggesting an origin purely from massive stars to be unlikely (though not impossible).

On the other hand, powerful broad H$\alpha$ emission is commonly seen in core collapse and/or super-luminous supernovae (e.g., \citealt{Gutierrez17}), and typically decays on timescales of several years at most. As a rudimentary test of a supernova origin to the broad H$\alpha$ component we detect in J0822$+$2241, we obtained updated Palomar/DoubleSpec spectroscopy of J0822$+$2241 $\sim$\,20 years ($\sim$\,16 years in rest-frame) after the original SDSS spectrum was taken. The corresponding comparison between the [S\textsc{ii}]\,$\lambda\lambda$6716,\,6731 doublet, the [N\textsc{ii}]\,$\lambda\lambda$6548,\,6583\,$+$\,H$\alpha$\,$\lambda$6563 complex and the H$\beta$\,$\lambda$4861 line are presented in Section~\ref{app1:palomar} and Figure~\ref{fig:gp2_palomar_comp}, in which the line profiles are remarkably similar across the $\sim$\,16 year baseline within expected calibration-based systematic uncertainties.

Thus based on the evidence we have in hand, it is likely that the broad H$\alpha$ line identified in J0822$+$2241 is powered by the broad line region surrounding an AGN. Future observations (e.g., time-resolved spectroscopy and/or resolved Integral Field Unit observations) would be required to definitively prove that the broad H$\alpha$ component is indeed AGN-powered.

\subsubsection{Estimating Extinction with the Balmer Decrement}\label{subsubsec:hbeta}

To access the intrinsic broad H$\alpha$\,$\lambda$6563 flux measured from the SDSS spectrum of J0822$+$2241, we additionally measure the line-of-sight extinction from the H$\beta$\,$\lambda$4861 emission line using the Balmer decrement. Since we find evidence for a significant broad component to the H$\alpha$ emission line, it is plausible a priori that a broad component to H$\beta$ exists. However, due to our use of a simple pseudo continuum model we cannot easily rely on the same [S\textsc{ii}] narrow line template to account for the narrow component to the H$\beta$ line, which would strictly require the same powerlaw to explain the continuum over $\gtrsim$\,1500\,${\rm \AA}$ between the [S\textsc{ii}] doublet and H$\beta$. Previous analyses of AGN candidates have focused on using the [O\textsc{iii}] emission lines to provide a narrow line template for H$\beta$ (e.g., \citealt{Liu25}). However, Green Pea galaxies are defined to have extremely bright and complex [O\textsc{iii}] emission, often including prominent broad components (e.g., \citealt{Izotov11}). Thus we purposefully avoid using [O\textsc{iii}] to interpret the narrow component of H$\beta$, and instead fit the H$\beta$ line by itself over the rest-frame wavelength range 4750\,--\,4950\,${\rm \AA}$.

\begin{figure}
\centering
\includegraphics[width=0.99\columnwidth]{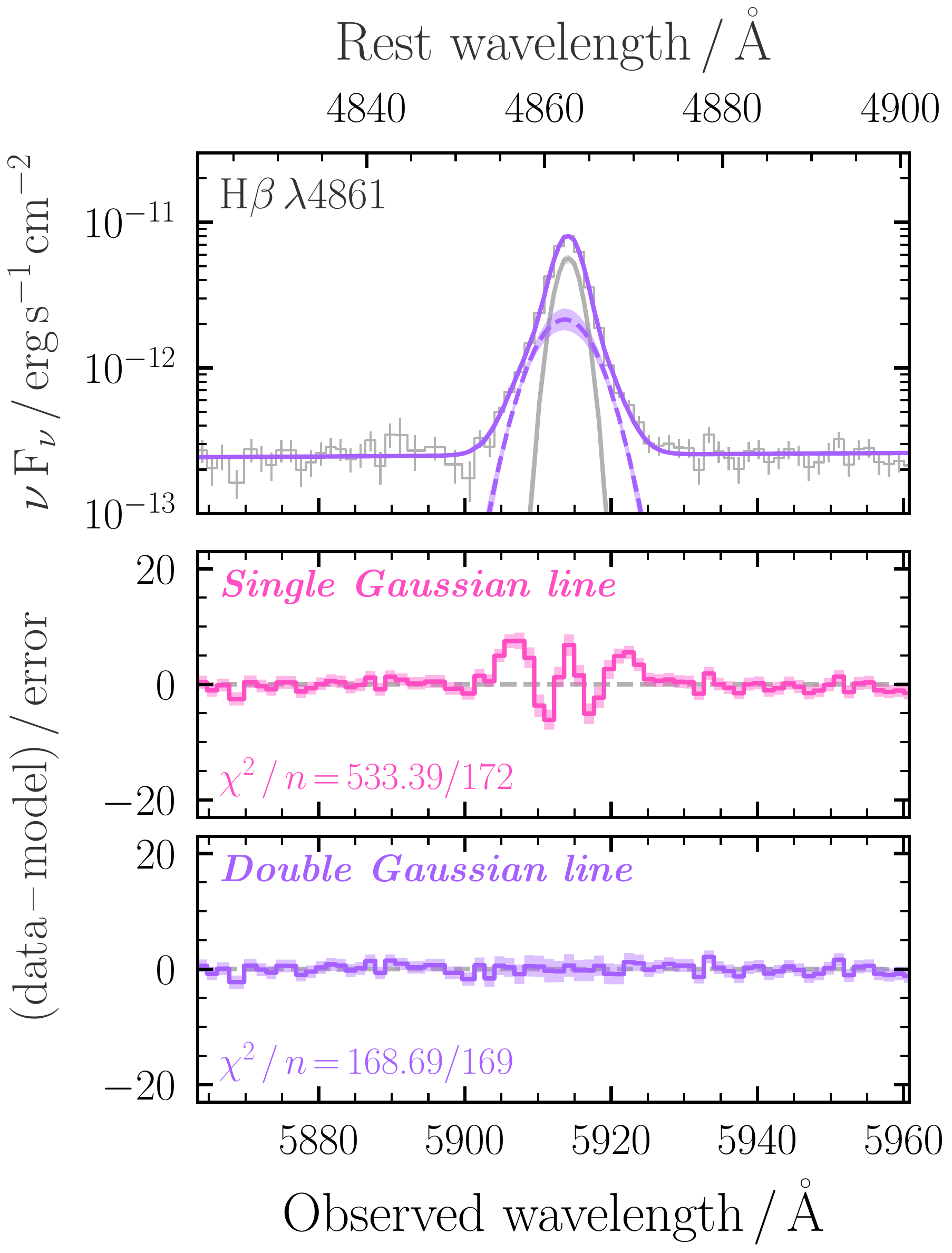}
\caption{(Top) SDSS H$\beta$\,$\lambda$4861 emission line of J0822$+$2241, with the spectral fit described in Section~\ref{subsubsec:halpha}. The solid purple line shows the total model, and the dashed purple line with shaded region shows the posterior constraint on the second Gaussian component to H$\beta$. The grey line and associated shading shows the posterior constraint on the first narrower Gaussian component to the H$\beta$ line. (Center) Residuals from an equivalent fit using a single Gaussian component. (Bottom) Residuals arising from the model with an additional Gaussian component to the H$\beta$ line included, relevant to the spectral fit shown in the top panel.}
\label{fig:gp2_hbeta}
\end{figure}

Figure~\ref{fig:gp2_hbeta} presents our phenomenological fitting to the H$\beta$ line. The fitting process is analogous to our spectral fitting of the [S\textsc{ii}] doublet in Section~\ref{subsubsec:sii}, in which we trial a single Gaussian model for the H$\beta$ line, followed by a double Gaussian model. For the single Gaussian line model, the continuum slope and normalisation as well as the line centroid, width and normalisation were allowed to vary giving five free parameters. For the double Gaussian model, the line centroid, width and normalisation of both components were allowed to vary freely, giving eight free parameters. We clearly find that the 1--Gaussian model is incapable of fitting the relatively broad base of the H$\beta$ emission line (c.f. centre panel of Figure~\ref{fig:gp2_hbeta}), giving a chi-squared of 533.39 with 172 degrees of freedom ($\chi^{2}_{n}$\,=\,3.10). Including the extra Gaussian component in the 2--Gaussian model substantially improves the fit, giving a chi-squared of 168.69 with 169 degrees of freedom ($\chi^{2}_{n}$\,=\,0.99). However, we find that the second (broader) Gaussian component has a width of FWHM$_{{\rm H}\beta,\,{\rm component}\,2}$\,=\,500\,$\pm$\,20\,km\,s$^{-1}$, which is insufficiently broad to have a high likelihood of being AGN-powered. Given the lack of a component comparably broad to the broad component of H$\alpha$, we make the conservative assumption that the total flux of both components comprising the H$\beta$ emission line constitutes the narrow H$\beta$ flux used to derive the Balmer decrement, which equates to log\,$F_{{\rm H}\beta,\,{\rm narrow,\,obs}}$\,=\,$-14.08$\,$\pm$\,0.01, after correcting for Milky Way extinction.

The corresponding Balmer decrement arising from the ratio between the narrow H$\alpha$ and H$\beta$ fluxes is $F_{{\rm H}\alpha,\,{\rm narrow,\,obs}}$\,/\,$F_{{\rm H}\beta,\,{\rm narrow,\,obs}}$\,=\,4.2\,$\pm$\,0.6. We translate the observed Balmer decrement into a nebular colour excess value of $E(B-V)$\,=\,0.3\,$\pm$\,0.1\,mag for J0822$+$2241 using Equation~(4) of \citet{Dominguez13} with the reddening law of \citet{Calzetti00} assuming an intrinsic H$\alpha$/H$\beta$ ratio of 2.86 \citep{Osterbrock89}. Again using \citet{Calzetti00}, we estimate the extinction at rest-frame H$\alpha$ to provide an intrinsic absorption-corrected broad H$\alpha$ luminosity of log\,$L_{{\rm H}\alpha,\,{\rm broad,\,int}}$\,/\,erg\,s$^{-1}$\,=\,42.1$^{+0.3}_{-0.2}$.

\subsection{Spectral Energy Distribution Analysis}\label{subsec:reslightning}

In the left panel of Figure~\ref{fig:gp2_sed}, we show the X-ray to infrared broadband SED for J0822$+$2241, along with model constraints from {\tt Lightning}. Our models prefer a solution in which both stellar and AGN processes are important in different regimes. For instance, stellar emission is expected to be important in the UV-to-optical regime and potentially dominate at wavelengths longer than $\sim$\,10\,$\mu$m due to cold dust emission. In contrast an AGN component is required to dominate the detected X-ray emission, with potentially important contributions to the UV-to-optical regime, and to dominate the $\sim$\,3\,--\,10\,$\mu$m near-to-mid infrared regime. It is useful to also note that the total infrared SED encompassed by the four-band \textit{WISE} photometry is characteristically steep, as previously identified by \citet{Kawamuro19}, but \texttt{Lightning} is able to predict that the red spectral shape is a result of AGN {\em and} host galaxy processes.

We find that the corresponding galaxy parameter posteriors derived with \texttt{Lightning} to be in basic agreement with those derived in the literature for J0822$+$2241 from the optical and UV portions of its SED. However, on average we find larger uncertainties due to the requirement for a joint contribution to the UV-to-optical regime from AGN and stellar processes. Specifically, we constrain the average star formation rate over the last 10\,Myr to be SFR$_{\,{\rm t}\,<\,10\,{\rm Myr}}$\,=\,44$^{+50}_{-38}$\,M$_{\odot}$\,yr$^{-1}$ and the integrated galaxy stellar mass to be $M_{*}$\,=\,$1.8^{+0.8}_{-1.5}$\,$\times$\,10$^{10}$\,M$_{\odot}$.
 
For the AGN component, the X-ray and near-to-mid infrared data are able to place some constraints on black hole mass using the parametrisation within the \texttt{qsosed} model, finding $M_{\rm BH}$\,=\,1.2$^{+2.5}_{-0.2}$\,$\times$\,10$^{6}$\,M$_{\odot}$. Combined with the constraint on stellar mass, \texttt{Lightning} predicts J0822$+$2241 to lie far closer to the stellar mass versus black hole mass relation derived for typical broad-line dwarf AGN from \citet{Reines15} (see Figure~\ref{fig:r15_relation}). As described in Section~\ref{sec:mstel}, the addition of an AGN component can drastically alter the predicted stellar mass in a non-obvious manner, meaning that the true value may lie anywhere between the value derived by \citet{Izotov11} and from our \texttt{Lightning} fit. The SED fitting additionally yields an AGN line-of-sight obscuring column density of log\,$N_{\rm H}/{\rm cm}^{-2} = 22.7^{+u}_{-0.1}$, which is significantly larger than that determined from the X-ray data alone.

As shown in the left side of the left panel of Figure~\ref{fig:gp2_sed}, the \texttt{Lightning} solution leaves an elevated residual in the X-ray band at $\lesssim$\,1\,keV. Some portion of the soft X-ray emission is modelled with stellar emission from X-ray binaries, which have a different absorption prescription to the AGN. At present the {\tt Lightning} models are expected to underpredict the X-ray emission in the soft X-ray region of the SED, as it does not include emission from hot gas, nor does it include a metallicity dependence to the X-ray binary component. Recent studies have shown that X-ray binary emission is significantly elevated in low-metallicity environments, and may plausibly be a factor of \hbox{$\approx$3--5} times higher than our model predictions for J0822$+$2241 due to its relatively low metallicity \citep[see, e.g.,][]{Lehmer24,Kyritsis2025}. Nonetheless, such a soft X-ray enhancement would still be insufficient to explain the broadband X-ray spectrum without a dominant contribution from an AGN, in agreement with our detailed X-ray fits of the \textit{XMM-Newton} data. We note that performing the above SED fitting using the X-ray data from epoch~1, instead of epoch~2, yields consistent results for SFR$_{\,{\rm t}\,<\,10\,{\rm Myr}}$, $M_{\rm BH}$, $M_{*}$, and $N_{\rm H}$, albeit with weaker constraints due to the lower signal-to-noise data.

\subsection{The UV-to-optical Spectral Shape}\label{subsubsec:v}

\citet{Lin25} recently showed that a subset of Green Pea galaxies hosting broad permitted Balmer lines have `V' shaped continua in $\nu$\,F$_{\nu}$ space in close analogy to Little Red Dots, following the spectral slope parametrisations of \citet{Kocevski24}. A remaining question from our broadband SED fitting of J0822$+$2241 is thus whether or not the underlying continuum shape in the UV-to-optical regime is consistent with that of Little Red Dots. To test this possibility, we consider the spectral slopes either side of the Balmer break in the rest-frame wavelength ranges of 1000\,--\,3645\,${\rm \AA}$ and 3645\,${\rm \AA}$\,--\,1\,$\mu$m for the UV and optical spectral slopes, respectively. All spectral slopes were derived by fitting a linear relation between logarithmic observed wavelength and observed flux density in AB magnitudes assuming $m_{\rm AB}$\,=\,$-2.5(\beta\,+\,2)$\,log\,$\lambda$\,$+$\,$c$, where $\beta$ is the spectral slope \citep{Kocevski24}. The linear relation was fit using \texttt{UltraNest}, assuming a uniform prior for the gradient from -10\,--\,10 and log-uniform priors for the y-intercept and intrinsic scatter in the y direction ranging from 0.1\,--\,100 and 0.001\,--\,1 respectively.

\begin{figure*}
\centering
\includegraphics[width=0.99\textwidth]{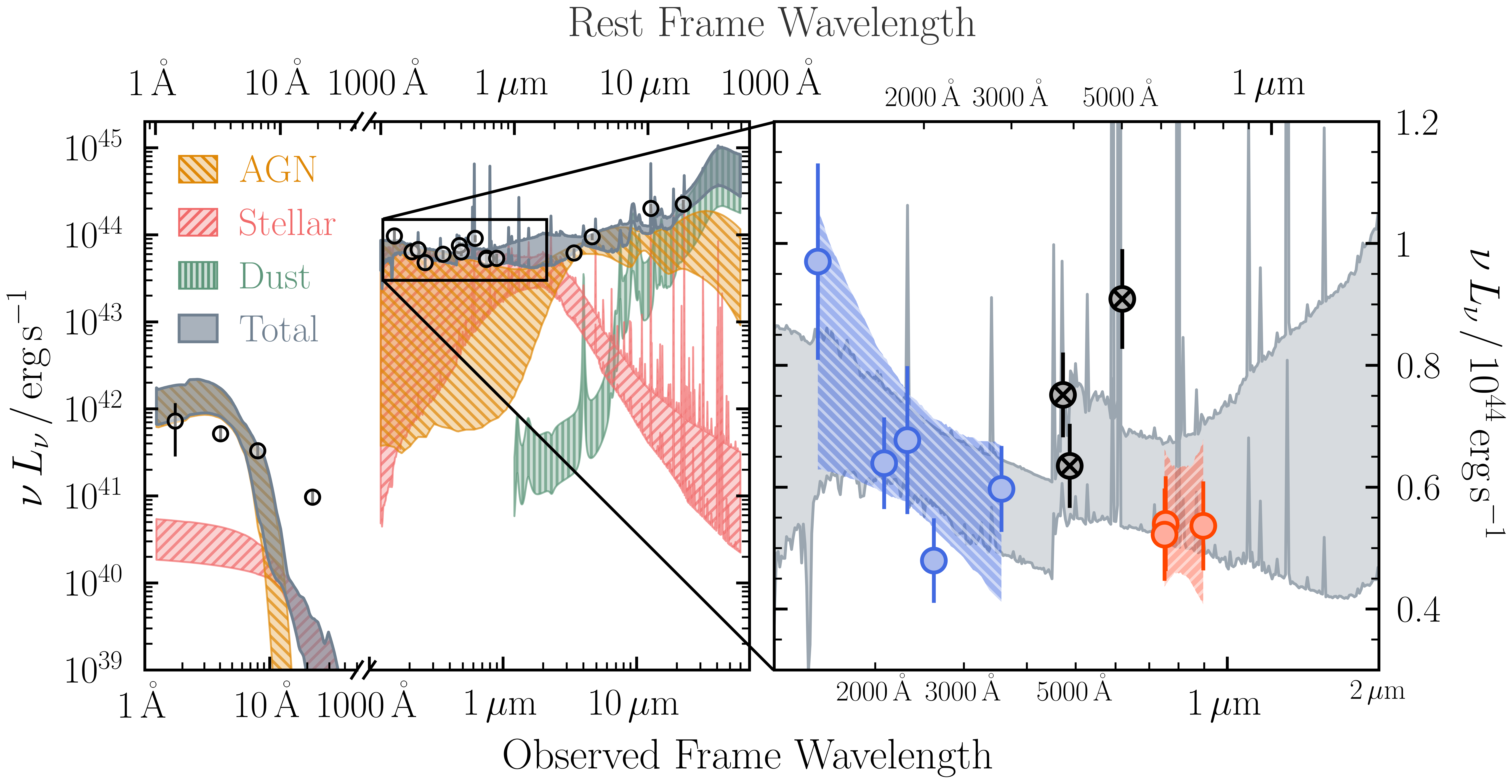}
\caption{(Left) Broadband SED of J0822$+$2241 (black circular points with associated 68\% errorbars), fit with the \texttt{Lightning} software package. The total model is shown with a dark grey shaded region, whereas the constituent AGN, attenuated stellar and dust components are shown with hatched golden, red and green regions, respectively. The left-hand portion of the plot shows the \texttt{Lightning} fit to the \textit{XMM-Newton} data of epoch~2. (Right) A zoom-in of the UV-to-near-infrared portion of the SED. By fitting slopes to the spectrum blueward and redward of the Balmer break, we constrain the overall UV and optical slopes, akin to \citet{Kocevski24}, to compare to the expected spectral shape from Little Red Dots. All photometry blueward of the Balmer break is used to derive the UV slope, given by the hatched blue region. Redward of the Balmer break, substantial emission line contamination from [O\textsc{ii}]\,$\lambda\lambda$3726,\,3729 and [O\textsc{iii}]\,$\lambda\lambda$4959,\,5007 is expected, such that we fit the optical slope between a restricted observed wavelength range of $\sim$\,7000\,--\,9000\,${\rm \AA}$. The corresponding optical spectral slope range constrained is shown with a red hatched region.}
\label{fig:gp2_sed}
\end{figure*}

The corresponding UV slope derived by fitting five photometry points blueward of the Balmer break (c.f. right panel of Figure~\ref{fig:gp2_sed}) is $\beta_{\rm UV}$\,=\,$-1.5^{+0.4}_{-0.7}$, which satisfies the \citet{Kocevski24} Little Red Dot UV slope requirement of $\beta_{\rm UV}$\,$<$\,$-0.37$. By comparing the optical photometry for J0822$+$2241 with the predicted \texttt{Lightning} fit, there is expected spectral line contamination from [O\textsc{ii}]\,$\lambda\lambda$3726,\,3729 and [O\textsc{iii}]\,$\lambda\lambda$4959,\,5007. We thus excluded the photometry associated with those lines from our spectral slope estimates. The corresponding slope measured between the three optical photometric points shown in Figure~\ref{fig:gp2_sed} is $\beta_{\rm optical}$\,=\,$-1.3^{+2.5}_{-2.2}$. Clearly, even though the slope measurement is consistent with the \citet{Kocevski24} requirement of $\beta_{\rm optical}$\,$>$\,0, the uncertainties are substantial due to the lack of reliably photometry to estimate the spectral slope from. The predicted \texttt{Lightning} model $>$\,1\,$\mu$m clearly increases on average to reproduce the red near-to-mid infrared \textit{WISE} photometry, but the observed optical spectrum appears to be inconsistent with Little Red Dots.

\section{Estimating the Black Hole and Host Galaxy Stellar Masses}\label{sec:mstel}
Two important parameters to consider for J0822$+$2241 in light of the presence of broad H$\alpha$ and our broadband SED fits are the black hole and host galaxy stellar masses. For example, previous work has found evidence suggesting compact low mass galaxies with broad permitted optical lines host `overmassive' central black holes, with black hole masses exceeding predictions from scaling relations depending on stellar mass (e.g., \citealt{Lin25,Juodzbalis25}). Such an overmassive scenario could hint towards previous episodes of prolonged rapid black hole growth. However, the stellar mass of luminous compact galaxies such as J0822$+$2241 is notoriously difficult to measure due in part to the requirement for a correct treatment of ionised gas emission as well as the general difficulty associated with characterising old stellar populations (see e.g., Section~5 of \citealt{Izotov11}). In total we consider three stellar mass measurements for J0822$+$2241 derived with different methods from \citet{Kauffmann03,Brinchmann04}, \citet{Cardamone09} and \citet{Izotov11}, yielding stellar masses of $M_{*}$\,=\,4\,$\times$\,10$^{9}$, 2\,$\times$\,10$^{9}$ and 3\,$\times$\,10$^{8}$\,M$_{\odot}$, respectively. The stellar mass of \citet{Kauffmann03,Brinchmann04} was measured by multiplying its dust- and K-corrected $z$-band luminosity by a $z$-band mass-to-light ratio estimation, whereas \citet{Cardamone09} measured stellar masses by convolving the observed SDSS spectral continuum with 19 medium band filters and fitting stellar population models with the observed \textit{GALEX} UV data. \citet{Cardamone09} note that due to the difficulty associated with properly accounting for older stellar populations, a minimum systematic uncertainty of 0.3\,dex should be considered. By considering the contribution from stellar and ionised gas emission, \citet{Izotov11} fit the full SDSS spectrum (continuum and lines) of J0822$+$2241 assuming a recent burst of star formation combined with a prior continuous and constant episode of star formation. The substantially smaller stellar mass than \citet{Kauffmann03}, \citet{Brinchmann04} and \citet{Cardamone09} derived is due to the extra contribution from gaseous continuum emission that is accounted for by \citet{Izotov11}.

\begin{figure}
\centering
\includegraphics[width=0.99\columnwidth]{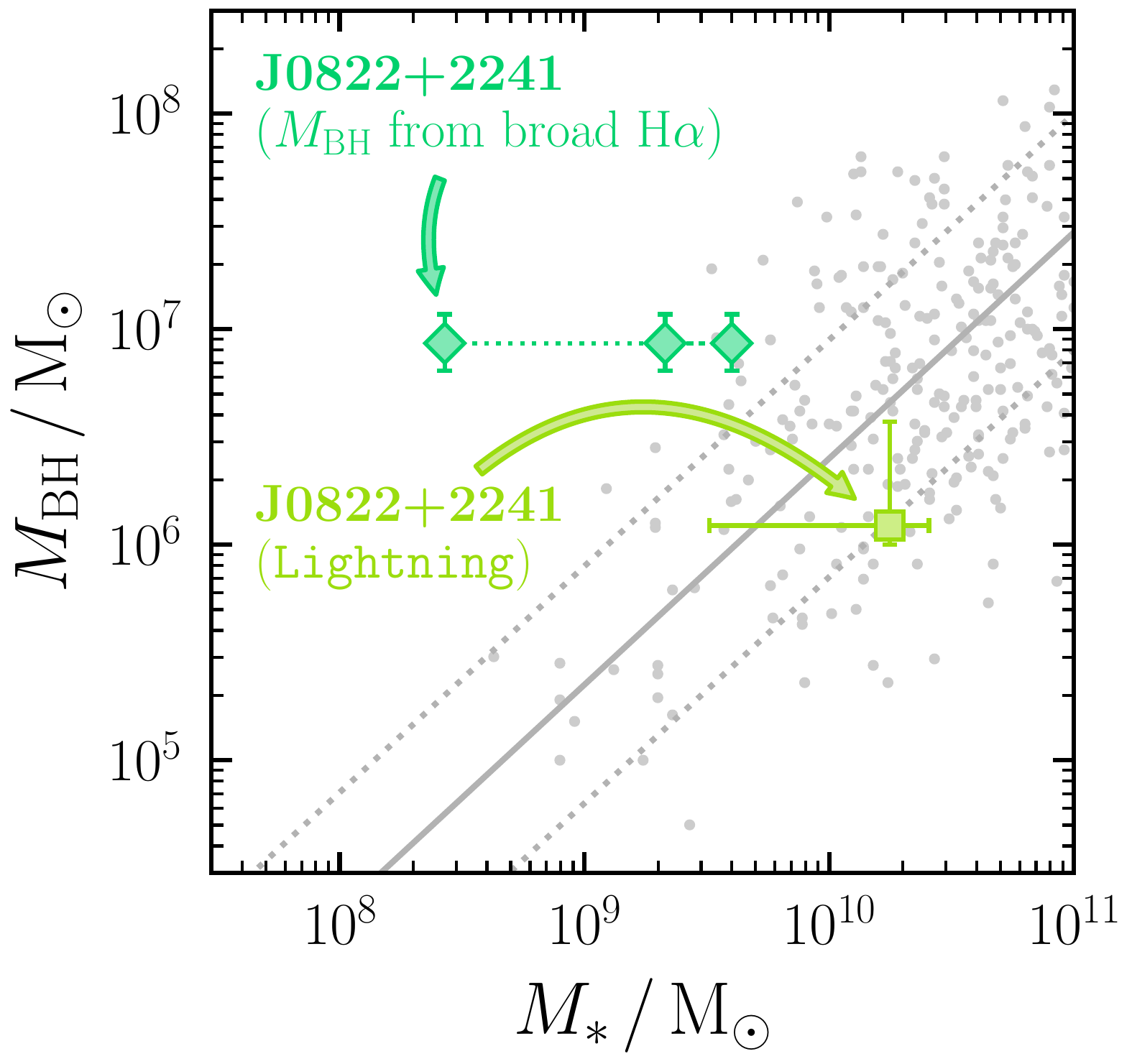}
\caption{The local broad line dwarf AGN sample with corresponding best-fit relation and intrinsic scatter between stellar mass and black hole mass from \citet{Reines15}. The black hole mass measured for J0822$+$2241 using the broad H$\alpha$ component identified in the SDSS optical spectrum (c.f. Section~\ref{subsec:obscuredagn}) is overplotted with a green diamond, together with three literature values of stellar mass taken from \citet{Izotov11}, \citet{Cardamone09}, \citet{Kauffmann03,Brinchmann04} in the left, center and right diamonds, respectively. The alternative black hole vs. stellar mass derived from \texttt{Lightning} that considers the presence of an AGN (c.f. Section~\ref{subsec:reslightning}) is shown with a lighter green square.}
\label{fig:r15_relation}
\end{figure}

To derive a black hole mass for J0822$+$2241, we first consider our absorption-corrected broad H$\alpha$ luminosity and FWHM from Section~\ref{subsubsec:hbeta} together with the black hole mass relation given in Equation~(5) of \citet{Reines13}. The corresponding black hole mass found for J0822$+$2241 from the broad H$\alpha$ line is M$_{{\rm BH,\,H}\alpha}$\,=\,8$_{-2}^{+3}$\,$\times$\,10$^{6}$\,M$_{\odot}$, though we note this mass prediction does fundamentally rely on an extrapolation from more luminous AGN that may not hold for low metallicity systems such as Green Pea galaxies (see Section~3.5 of \citealt{Reines13} for further discussion). Figure~\ref{fig:r15_relation} presents the H$\alpha$-based black hole mass for J0822$+$2241 together with all three stellar masses over-plotted on the black hole mass versus stellar mass scaling relation of \citet{Reines15}.

On first interpretation, the H$\alpha$-based black hole mass estimation found for J0822$+$2241 lies significantly outside the intrinsic scatter of \citet{Reines15} for all literature stellar masses, though the most extreme offset is found with the stellar mass of \citet{Izotov11}. However, all literature methods selected do not consider an additional contribution from an AGN. If the observed optical continuum of J0822$+$2241 were also found to include a contribution from an AGN (as suggested by our \texttt{Lightning} SED fits), the stellar masses derived could have significant uncertainties that are not shown in Figure~\ref{fig:r15_relation} (see \citealt{Buchner24} for a detailed discussion of AGN-induced stellar mass uncertainties). On one hand, if the AGN emission were to make the object brighter in all bands without adding significant colour terms, the current stellar masses could be considered upper limits. A smaller stellar mass would then make J0822$+$2241 more of an outlier compared to the \citet{Reines15} relation. On the other hand, if the contribution from an AGN were to make the observed continuum bluer without significantly affecting the total flux in the redder bands, the contribution from younger stellar populations could be over-estimated. As younger stellar populations have a lower mass-to-light ratio, the stellar masses could then be considered as lower limits.

For completeness, we additionally plot the self-consistently derived black hole and stellar masses derived from \texttt{Lightning}. Though the uncertainties are larger, the masses derived with \texttt{Lightning} are consistent with the stellar mass versus black hole mass relation of \citet{Reines15}. Our \texttt{Lightning} SED fits thus suggest that there is a plausible parameterisation that includes an AGN in which the black hole mass is not overmassive relative to the stellar mass scaling relation of \citet{Reines15}. We thus conclude that the current evidence is insufficient to make any strong claims that depend on the stellar mass of J0822$+$2241.

\section{Discussion}\label{sec:discussion}

Given the observed X-ray, optical and broadband SED properties of J0822$+$2241, we now consider three physical scenarios to explain the primary X-ray power source.

\subsection{Obscured X-ray Emission from an AGN}\label{subsec:obscuredagn}
A natural explanation for a persistent 2\,--\,10\,keV luminosity of $\sim$\,10$^{42}$\,erg\,s$^{-1}$ combined with decreasing soft X-ray luminosity over a baseline of 6.2\,years is from obscuration changes surrounding a Seyfert-like AGN. The vast majority of AGN are known to be obscured (e.g., \citealt{Ueda14,Buchner15,Ricci17_bassV,TorresAlba21,Tanimoto22,Boorman25_nulands}), and soft X-ray signatures of obscuration changes have been confirmed in numerous previous studies of the local Seyfert population (e.g., \citealt{Risaliti02,Rivers11,Walton14,Markowitz14,Lefkir23,TorresAlba23,Pizzetti25}). Specifically in the case of J0822$+$2241, changing photoelectric absorption could deplete the 0.5\,--\,2\,keV flux between epochs~1 and~2 without significantly altering the harder 2\,--\,10\,keV flux, resulting in a spectral hardening akin to that found in Section~\ref{sec:results}.

\begin{figure*}
\centering
\includegraphics[width=0.99\textwidth]{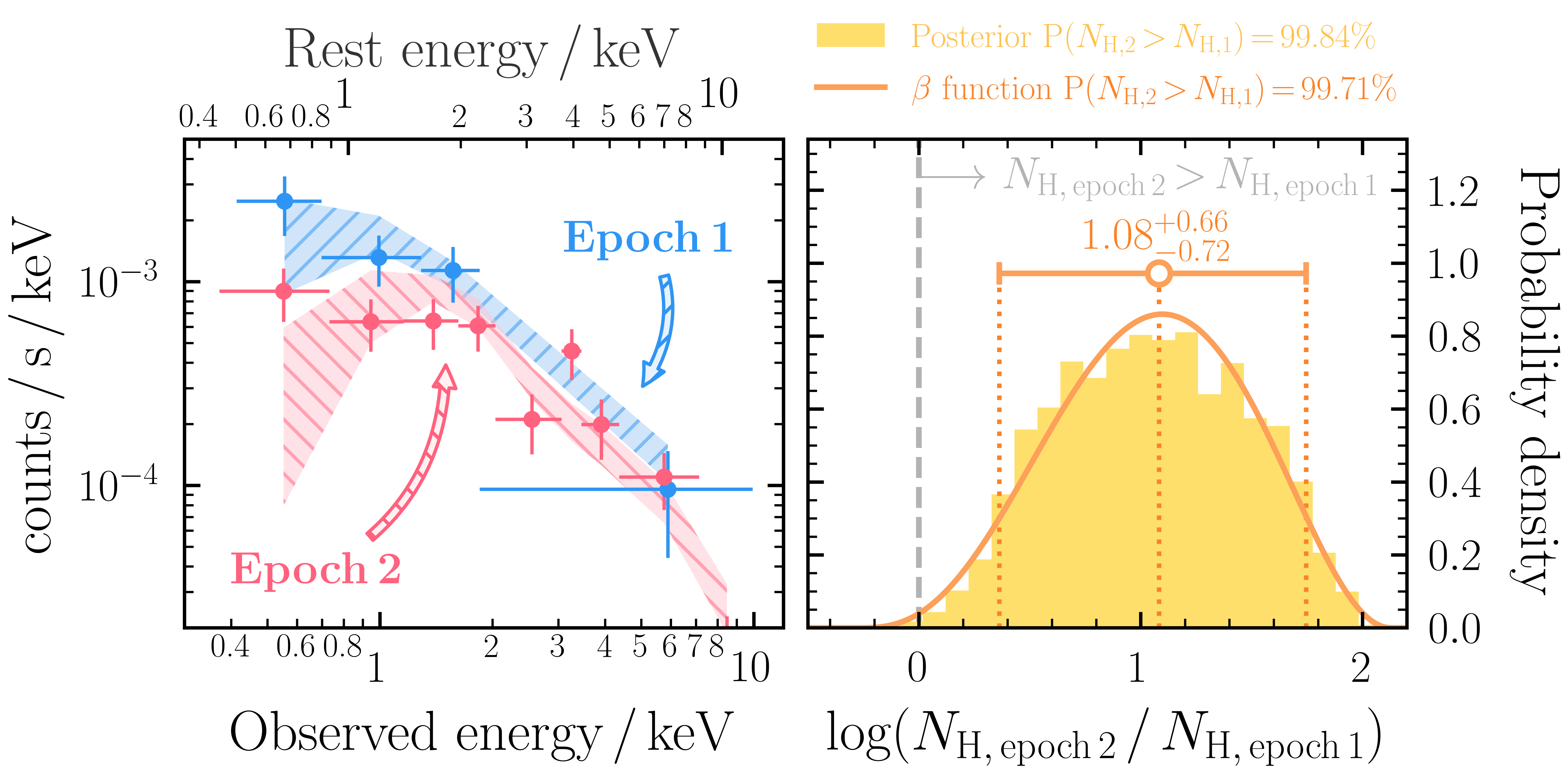}
\caption{(Left) Simultaneous X-ray spectral fit to the two epochs of data from J0822$+$2241 using a redshifted power law with variable line-of-sight obscuration between epochs. (Right) Posterior distribution for the logarithmic ratio of column density from epoch~2 to~1. All values above zero are consistent with the column density increasing from epoch~1 to~2. The solid dark orange curve shows an analytical beta function fit to the posterior distribution. Both the raw posterior and beta function agree that an increase in column density is required to $>$\,99.7\% probability.}
\label{fig:gp2_nhvar}
\end{figure*}

We fit an absorbed powerlaw (\texttt{zTBabs*cabs*zpowerlw} in Xspec parlance) to both epochs simultaneously whilst allowing the line-of-sight column density to vary independently between epochs. The resulting spectral fits with folded posterior ranges are shown in the left panel of Figure~\ref{fig:gp2_nhvar}. Both epochs are explained well by the model, resulting in an observed-frame 2\,--\,10\,keV flux of log\,$F_{2-10\,{\rm keV}}$\,/\,erg\,s$^{-1}$\,cm$^{-2}$\,=\,$-14.0$\,$\pm$\,0.1 and intrinsic photon index of $\Gamma$\,=\,1.7$^{+0.1}_{-0.2}$, fully consistent with the typical photon index values found from surveys of AGN (e.g., \citealt{Ricci17_bassV}). We find line-of-sight column densities of log\,$N_{\rm H}$\,/cm$^{-2}$\,=\,20.7$_{-u}^{+0.7}$ and 21.8\,$\pm$\,0.3 for epochs~1 and~2, respectively. Our use of BXA enables us to propagate the posteriors of column density per epoch into a single posterior on the column density ratio between each epochs which, by design, incorporates all covariance associated with the fit. By integrating the probability mass encompassed in the epoch~2-to-epoch~1 column density ratio posterior with values above unity, we find an increase in obscuration between 2013 and 2020 is required to $>$\,99.8\% confidence. In case integrating the posterior derived from UltraNest is affected by discrete sampling effects, we additionally fit the column density ratio posterior with a flexible beta function following the method of \citet{Baronchelli20}. The corresponding one dimensional posterior and beta function fit are shown in the right panel of Figure~\ref{fig:gp2_nhvar}, in which the probability of an increase in obscuration between 2013 and 2020 is required to $>$\,99.7\% confidence. The beta function fit additionally yields a fractional column density increase of log($N_{\rm H,\,2}$\,/\,$N_{\rm H,\,1}$)\,=\,1.08$^{+0.66}_{-0.72}$. We note that since the column density in epoch~1 is consistent with being below the Galactic value (i.e. unconstrained at the lower end), our column density ratio between epochs~2 and~1, as well as the corresponding probability for an increase in column density can both conservatively be considered lower limits.

To constrain the Eddington ratio in the obscured AGN scenario, we first estimate the bolometric luminosity using the relation between bolometric correction and 2\,--\,10\,keV luminosity for AGN presented by \citet{Duras20}. Combined with the 2\,--\,10\,keV luminosity constrained for J0822$+$2241, we find a bolometric correction of $\kappa_{\rm X}$\,=\,15.5\,$\pm$\,0.1 which together give a bolometric luminosity of $L_{\rm bol}$\,/\,erg\,s$^{-1}$\,=\,43.2\,$\pm$\,0.1. By combining the H$\alpha$-derived black hole mass and bolometric luminosity estimates, we compute an Eddington ratio for J0822$+$2241 of $\lambda_{\rm Edd}$\,=\,1.4$_{-0.7}^{+0.4}$\,\% (log\,$\lambda_{\rm Edd}$\,=\,$-1.8^{+0.1}_{-0.2}$). Interestingly, the Eddington ratio range constrained is consistent with the effective Eddington limit on dusty gas (c.f. \citealt{Fabian08,Ricci17_effedd}). Previous work has shown the effective Eddington limit on dusty gas to coincide with a sharp decline in the fraction of obscured AGN selected within the 70-month \textit{Neil Gehrels Swift}/BAT survey \citep{Ricci17_effedd,Ananna22,Ricci22,Ricci23}, consistent with a radiation pressure dependent covering factor of (sub-Compton-thick) material. If the same principle were to apply for J0822$+$2241, the obscuration change observed between epoch~1 and~2 could have arisen from an outflowing configuration.

\subsection{Unobscured X-ray Emission from an AGN}\label{subsec:unobscuredagn}

Soft X-ray variability is commonly observed in unobscured AGN (e.g., \citealt{Kara25} and references therein), though the properties observed for J0822$+$2241 are largely disparate from the broader unobscured population. For example, a possibility to consider for the soft X-ray emission of J0822$+$2241 is a variable soft excess component that is often observed in unobscured AGN (e.g., \citealt{Gierlinski04,Crummy06,Done12,Walton13,Waddell24,MadathilPottayil24}). In such a scenario the intrinsic photon index of the AGN coronal continuum would be harder than observed due to the contribution of the soft excess flux at $\lesssim$\,2\,keV. The most conservative scenario in producing the softest possible intrinsic photon index for J0822$+$2241 would be for the epoch~2 spectrum to feature no detectable soft excess. However, the observed photon index of the spectrum in epoch~2 ($\Gamma$\,=\,0.8$^{+0.2}_{-0.3}$) is already considerably harder than is typically observed for unobscured AGN that are found to host strong soft excesses (e.g., \citealt{Jin12a}), even before accounting for any potential soft excess in the fits to J0822$+$2241.

An alternative scenario to explain the power source of J0822$+$2241 is variable accretion onto an unobscured intermediate mass black hole that has low enough mass for the accretion disk spectrum to be detectable by \textit{XMM-Newton}. The observed photon index of J0822$+$2241 in epoch~2 ($\Gamma$\,=\,0.8$^{+0.2}_{-0.3}$) is already significantly harder than the photon index found in one of the hardest states of ESO\,243--49 HLX-1 ($\Gamma$\,=\,1.6\,$\pm$\,0.4; \citealt{Servillat11}) which is widely regarded as a strong intermediate mass black hole candidate. We therefore defined the minimum 0.5\,--\,2\,keV flux of an accretion disk component to be the difference in flux between both epochs. Assuming upper limits for detectable accretion disk emission in the observed ultraviolet (as measured by the \textit{XMM-Newton}/Optical Monitor UVW1 filter; \citealt{Page12}) and 2\,--\,10\,keV bands, we simulated accretion disk spectra with the \texttt{KYNbb} model \citep{Dovciak08,Dovciak04} whilst simultaneously varying black hole mass, spin, inclination, inner disk radius and accretion rate. We find a black hole mass of $M_{\rm BH}$\,=\,$1.1_{-0.9}^{+36.0}$\,$\times$\,10$^{4}$\,M$_{\odot}$ (16th, 50th and 84th percentiles), which is broadly consistent with the expectations from the stellar mass versus black hole mass scaling relation of \citet{Reines15} plotted in Figure~\ref{fig:r15_relation}. However, the black hole mass is significantly lower than the value derived using the broad H$\alpha$ emission line (c.f. Section~\ref{subsec:optres}). Furthermore, the accretion rate would need to be super Eddington, with $\dot{m}$\,$>$\,$\dot{m}_{\rm Edd}$ which in the literature to-date has tended to be associated with substantially steeper spectra than the $\Gamma$\,=\,0.8$_{-0.3}^{+0.2}$ observed in epoch~2 for J0822$+$2241. The properties of accreting compact objects at such large accretion rates are also not yet fully understood, but are likely associated with powerful accretion disc winds that would render our use of \texttt{KYNbb} insufficient to constrain the black hole mass in the first place \citep{King23}. Though undoubtedly an exciting possibility, we thus consider an unobscured intermediate mass black hole revealed from X-ray spectral variability for J0822$+$2241 to be unlikely.

\subsection{Off-Nuclear Contaminants}\label{subsec:ulxhlx}
As discussed in Section~\ref{sec:intro}, \citet{Adamcova24} has shown previously that J0822$+$2241 could not have a contribution of $>$\,20\% to the total observed 0.5\,--\,8\,keV luminosity from unresolved populations of X-ray Binaries. Due to the consistent 2\,--\,10\,keV luminosity between the two epochs, an individual short-term transient event producing the observed X-ray spectra such as a supernova is also effectively ruled out (e.g., \citealt{Dwarkadas14}). In addition, more X-ray-luminous supernovae are often accompanied by more rapid X-ray flux declines \citep{Dwarkadas12}. Thus the cumulative effect from a sufficiently high frequency of supernovae to appear as a constant X-ray flux over $\sim$\,6 years in rest-frame is additionally infeasible.

\begin{figure*}
\centering
\includegraphics[width=0.99\textwidth]{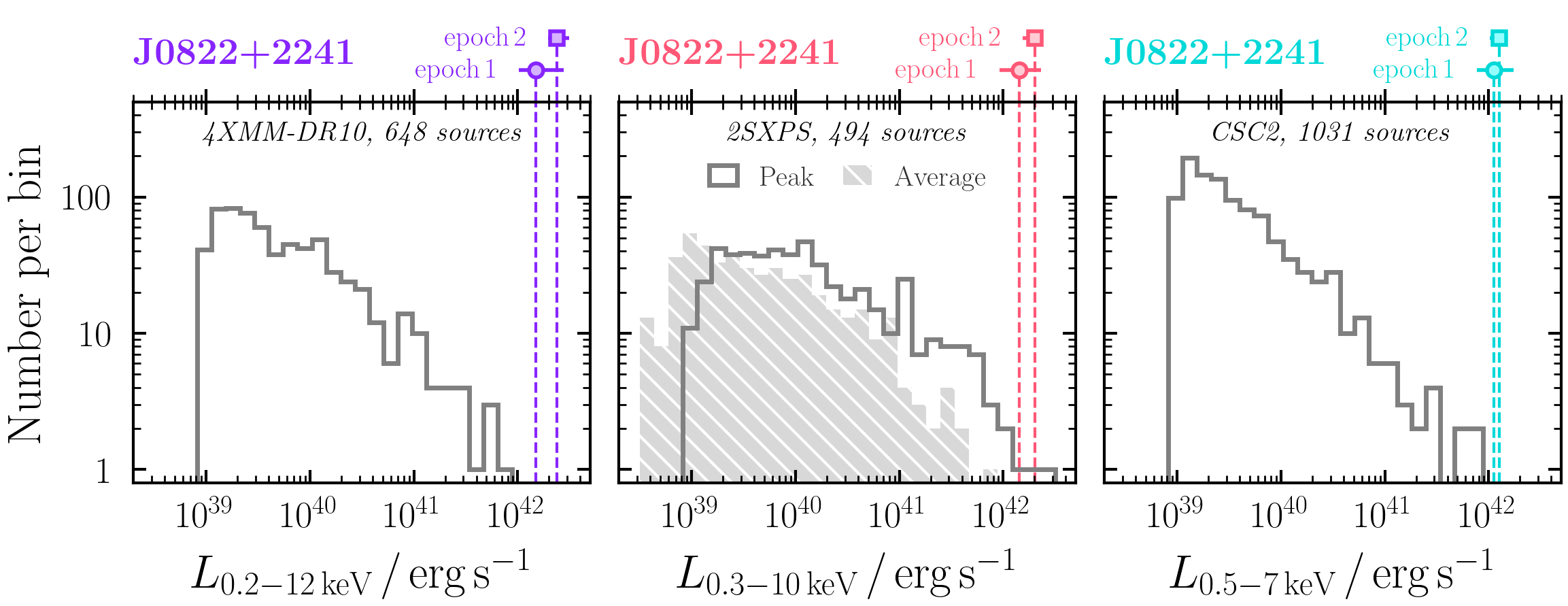}
\caption{Each panel shows the observed luminosities of J0822$+$2241 measured with a \texttt{diskbb}\,$+$\,\texttt{diskpbb} model in different bands, compared to the distributions of peak luminosity found by \citet{Walton22} for the ULX population in each of 4XMM-DR10 (left), 2SXPS (center) and CSC2 (right). Since luminosities are reported for different default passbands per instrument, we note that the J0822$+$2241 luminosities plotted per panel are measured for different passbands to allow a direct comparison. Clearly J0822$+$2241 is a significant outlier, with only one source being consistent with the epoch 2 luminosity measurement of J0822$+$2241 in the 2SXPS peak luminosity distribution.}\label{fig:gp2_ulx}
\end{figure*}

Owing to the small physical size of J0822$+$2241 with a NUV half light radius of 680\,pc \citep{Yang17}, the host galaxy is unresolvable by \textit{XMM-Newton}. Thus a remaining possibility to consider is that the X-ray emission we see from J0822$+$2241 arises from an unresolved number of Ultra-Luminous X-ray sources (ULXs, $L_{\rm X}$\,$>$\,10$^{39}$\,erg\,s$^{-1}$ -  see \citealt{King2023_ULX} for a review) and/or Hyper-Luminous X-ray sources (HLXs, $L_{\rm X}$\,$>$\,10$^{41}$\,erg\,s$^{-1}$; \citealt{Gao03}). Figure~\ref{fig:gp2_ulx} shows the observed X-ray luminosity of J0822$+$2241 in three bands from the individual epoch fitting with a powerlaw detailed in Section~\ref{sec:results}. Each measured rest-frame observed luminosity is compared to the peak luminosities seen in the multi-mission catalogue of ULXs derived by \citet{Walton22} from the fourth \textit{XMM-Newton} serendipitous source catalogue (4XMM-DR10; \citealt{Webb20}), the \textit{Neil Gehrels Swift} X-Ray Telescope Point-source Catalog (2SXPS; \citealt{Evans20b}) and the \textit{Chandra} Source Catalog DR2 (CSC2.0; \citealt{Evans20a}) in the left, center and right panels, respectively. The luminosities observed for J0822$+$2241 in both epochs are clearly extreme for individual members of the ULX population. Even though there is agreement with a small handful of sources identified to have peak luminosities in the 2SXPS survey consistent with J0822$+$2241, we note that the consistent 2\,--\,10\,keV luminosity over 6.2\,years suggests that the measured luminosities are not peak luminosities in the first place.

Though just contained within the populations plotted in Figure~\ref{fig:gp2_ulx}, current confirmed HLXs are considerably rarer \citep{MacKenzie23}. As discussed in Section~\ref{subsec:unobscuredagn}, a HLX powered by accretion onto an intermediate mass black hole (comparable to HLX-1) is unlikely to explain the observed variability and X-ray spectral shape of J0822$+$2241. We do also note that for a compact low mass galaxy such as J0822$+$2241, an individual accreting intermediate mass black hole as bright as $L_{\rm X}$\,$\sim$\,10$^{42}$\,erg\,s$^{-1}$ would not be a `contaminant' with regard to the AGN scenario. However, a remaining question is whether or not a neutron star-powered HLX could explain the observed properties of J0822$+$2241. The current brightest observed peak luminosity from a confirmed neutron star HLX is for NGC\,5097 ULX1 that is known to reach $L_{\rm X}$\,$\sim$\,10$^{41}$\,erg\,s$^{-1}$ \citep{Fuerst17,Israel17}, far lower than the luminosity found for J0822$+$2241. An interesting candidate HLX identified by \citet{Walton22,MacKenzie23} could be associated with IC\,1633, which corresponds to the only source in Figure~\ref{fig:gp2_ulx} that is consistent with the epoch~2 luminosity measurement of J0822$+$2241 within the 2SXPS. However, as noted by \citet{MacKenzie23}, that source lies in a galaxy with strong diffuse X-ray emission that may contaminate the soft X-ray flux measurements by \textit{Swift} and \textit{XMM-Newton}. Regardless, if the current known neutron star-powered HLX population are applicable to Green Pea galaxies, multiple sources would still be required to explain the X-ray spectral properties of J0822$+$2241. If on the other hand the X-ray flux from J0822$+$2241 were powered by the brightest neutron star HLX currently known, we re-iterate that the source would need to reach 2\,--\,10\,keV luminosities of $\gtrsim$\,10$^{42}$\,erg\,s$^{-1}$ on at least two separate occasions during a 6.2\,year baseline.

Our final consideration is thus whether a contribution of multiple U/HLXs within J0822$+$2241 could plausibly explain its observed X-ray properties. Such sources are known to exhibit a wide array of spectral variability \citep{Middleton15}. Thus at face value it seems possible that a population of unresolved (and causally disconnected) U/HLXs could produce a 2\,--\,10\,keV luminosity exceeding 10$^{42}$\,erg\,s$^{-1}$ with a combined variability pattern resembling the observed hardening of the \textit{XMM-Newton} spectrum of J0822$+$2241 from epoch~1 to epoch~2. \citet{Sutton12} showed with high-angular resolution \textit{Chandra} observations that a seemingly bright off-nuclear X-ray source detected by \textit{XMM-Newton} in the starburst galaxy NGC\,2276 was in fact several unresolved ULXs. However, for the spectrum of J0822$+$2241 to be explained by ULXs alone would require 10\,$<$\,$N_{\rm ULX}$\,$<$\,1000 individual sources with X-ray luminosities of $>$\,10$^{39}$\,$<$\,$L_{\rm X}$\,$<$\,10$^{41}$\,erg\,s$^{-1}$ at the times of the epoch~1 and~2 \textit{XMM-Newton} observations. Substantially higher numbers than this would also be needed if accounting for the expected transient nature of ULXs (e.g., \citealt{Brightman23}). To be comprised of multiple HLXs would require $\lesssim$\,10 individual sources in J0822$+$2241. Given that HLXs as a whole currently constitute $\sim$\,2\,--\,4\% of the entire detected ULX population \citep{Walton22,MacKenzie23}, $>$\,1 HLX in an individual galaxy is additionally unlikely.

Detailed studies of ULX populations in local galaxies have found a positive correlation between star formation rate and the number of ULXs detected, as well as a tentative enhanced star formation rate-normalised ULX rate at lower metallicities (e.g., \citealt{Mapelli10,Swartz11}). Considering the relation between star formation rate and ULX number from \citet{Mapelli10}, we would expect $\sim$\,20$^{+50}_{-10}$ ULXs to be present in J0822$+$2241 if using the star formation rate of 37\,$\pm$\,4\,M$_{\odot}$\,yr$^{-1}$ calculated by \citet[with uncertainty quoted at the 1$\sigma$ level]{Cardamone09}, with a slight potential enhancement arising from its low metallicity. Such high numbers of ULXs in a single galaxy are extremely rare in the local universe. One of the most extreme examples is that of the Cartwheel ring galaxy with at least 29 ULXs associated \citep{Salvaggio23}. However, conservatively including any off-nuclear sources studied by \citet{Salvaggio23}, the total average unabsorbed 0.5\,--\,10\,keV luminosity from potential ULXs in the Cartwheel galaxy would be $L_{0.5-10\,{\rm keV}}$\,$\sim$\,2\,$\times$\,10$^{41}$\,erg\,s$^{-1}$ \citep{Salvaggio23}, which is still a factor of five lower than the X-ray luminosity we find for J0822$+$2241. Furthermore, given existing ULX luminosity functions (e.g., \citealt{Luangtip15}), the expected number of high luminosity U/HLXs contained within a population of $\sim$\,20 U/HLXs would still be insufficient to explain the X-ray luminosity of J0822$+$2241.

A final test for the presence of U/HLXs in J0822$+$2241 that we consider is from radio observations. \citet{Borkar24} presented Very Large Array 6\,GHz and 10\,GHz observations of J0822$+$2241, finding an individual radio point source on $\sim$\,0.6\,--\,1\,arcsec scales ($\sim$\,3\,--\,5\,kpc at the redshift of J0822$+$2241) coincident with the center of the galaxy. The general lack of off-center point sources and/or extension with the radio measurements already suggests that powerful off-nuclear contaminants are not present. Using the radio spectral index derived by \citeauthor{Borkar24}, we find rest-frame 5\,GHz-to-2\,--\,10\,keV luminosity ratios of log\,$R_{\rm X}$\,=\,-2.8\,$\pm$\,0.3 and -3.0\,$\pm$\,0.1 for epochs~1 and~2, respectively. In comparison to the log\,$R_{\rm X}$ values measured by \citet{Terashima03}, such values are entirely consistent with those expected from low-luminosity and Seyfert-like AGN. In addition, \citet{Mezcua13,Mezcua15} show that X-ray binaries are expected to have log\,$R_{\rm X}$\,$<$\,-5.3, in contrast to low-luminosity AGN with -3.8\,$<$\,log\,$R_{\rm X}$\,$<$\,-2.8, the latter of which is consistent with J0822$+$2241 (see also \citealt{Argo18,Foord24}). Thus the joint X-ray and radio properties of J0822$+$2241 show that its high X-ray luminosity can be accounted for without a dominant population of U/HLXs at radio nor X-ray wavelengths.

Interestingly, \citet{Borkar24} show that the radio fluxes detected for J0822$+$2241 are consistent with the expectations from numerous star formation rate relations in the literature if predicted from the star formation rate of \citet{Cardamone09}. However, this star formation rate was predicted with spectral fitting of the H$\alpha$ line, for which our optical spectral fitting in Section~\ref{subsec:optres} now suggests an additional broad component is required. The presence of an AGN is known to often significantly bias estimates of star formation rate (e.g., \citealt{Kouroumpatzakis21} and references therein). For the case of J0822$+$2241, the broad component to H$\alpha$ could indicate the previous star formation rate is over-estimated. We additionally note that our independent star formation rate measurement with \texttt{Lightning} in Section~\ref{subsec:reslightning} is consistent with \citet{Cardamone09} though with substantially wider uncertainties that reach $\sim$\,6\,M$_{\odot}$\,yr$^{-1}$ within 90\% confidence. A lower star formation rate would yield a lower predicted star formation-powered radio flux, which could then reveal an AGN-powered radio excess. The overall agreement between the radio-to-X-ray ratio of J0822$+$2241 and other radio-loud local AGN suggests a non-negligible contribution to the radio emission from an AGN, which additionally supports the star formation rate of J0822$+$2241 being over-estimated.

\section{Relevance to JWST-detected AGN}\label{sec:jwstcomp}
\citet{Maiolino25} recently presented a compilation of intermediate-to-high redshift \textit{JWST}-detected AGN with no detected X-ray counterparts in some of the deepest \textit{Chandra} fields ever observed (see also \citealt{Ananna24,Yue24}). Though some of these undetected X-ray sources were the so-called elusive Little Red Dots \citep{Akins23,Kocevski24}, it is important to note that the X-ray non-detections are far more widespread than just this population, with almost every \textit{JWST}-detected AGN in the compilation of \citet{Maiolino25} lacking an X-ray counterpart. \citet{Maiolino25} poses two possible solutions for the lack of X-ray detections: Compton-thick dust-free gas obscuration or intrinsically soft X-ray spectra akin to Narrow Line Seyfert~1 galaxies.

\begin{figure}
\centering
\includegraphics[width=0.99\columnwidth]{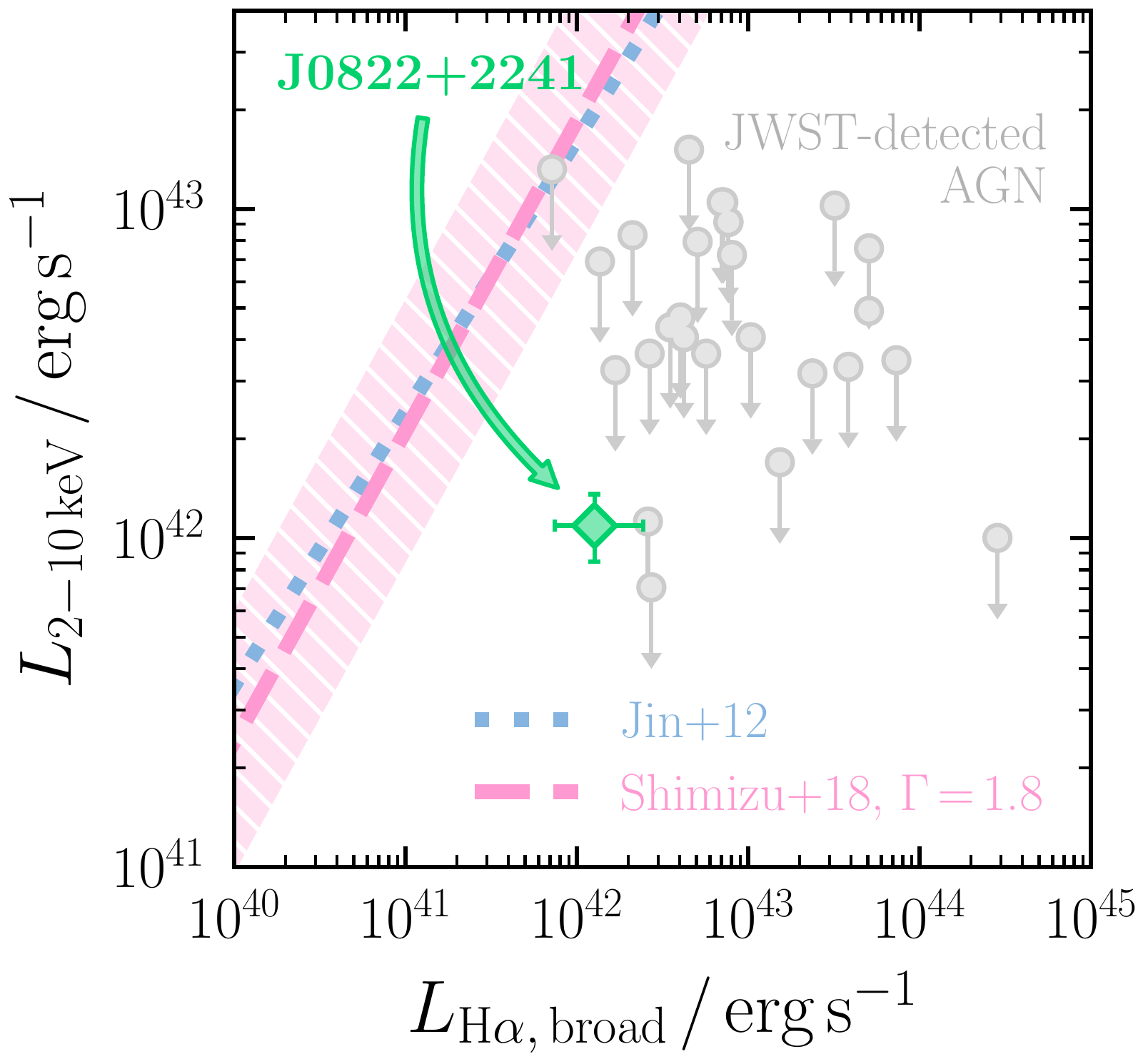}
\caption{Comparison between broad H$\alpha$ luminosity and 2\,--\,10\,keV luminosity measured for J0822$+$2241 and for the recent compilation of type~1 \textit{JWST}-detected AGN from \citet{Maiolino25}. We additionally plot two relations between broad H$\alpha$ luminosity and X-ray luminosity derived from samples in the local universe. First, we show with a pink dashed line (and associated shaded intrinsic scatter) the relation of \citet{Shimizu18}, constrained for type~1\,--\,1.2 AGN from the 70-month BAT catalogue. The 2\,--\,10\,keV luminosity for the relation was approximated from 14\,--\,195\,keV by assuming a powerlaw with photon index 1.8. For consistency, we additionally show the relation between narrow H$\alpha$-subtracted luminosity and 2\,--\,10\,keV luminosity from \citet{Jin12b} as a dotted blue line.}\label{fig:gp2_jwst_agn}
\end{figure}

The broad H$\alpha$ FWHM we find for J0822$+$2241 is fully consistent with the broad permitted line widths reported by \citet{Brooks24,Maiolino25} for the type~1 \textit{JWST}-detected AGN candidates displaying broad permitted lines. However, the equivalent width we measure is significantly below the median value found for the \textit{JWST} AGN of 570\,${\rm \AA}$. The comparably low width constrained for the broad component of H$\beta$ is consistent with some \textit{JWST} broad H$\alpha$ AGN candidates in which no significant broad H$\beta$ component  is detected \citep{Brooks24}. However, our measured X-ray properties for J0822$+$2241 do not agree with either scenario posed by \citet{Maiolino25} to explain the \textit{JWST} AGN. We find minimal obscuration for J0822$+$2241 and a standard unobscured AGN intrinsic photon index of $\Gamma$\,=\,1.7$^{+0.1}_{-0.2}$, far flatter than the steep shapes required to substantially deplete the observed X-ray flux from $z$\,$\gtrsim$\,6 sources with \textit{Chandra} nor the typical photon indices of more local Narrow Line Seyfert~1 AGN (e.g., \citealt{Jin12a}). An obvious first question to ask then is: how does J0822$+$2241 compare to the \textit{JWST}-detected AGN in terms of measured broad permitted H$\alpha$ and X-ray constraints? Figure~\ref{fig:gp2_jwst_agn} plots the broad H$\alpha$ luminosity versus 2\,--\,10\,keV X-ray luminosity for the type~1 AGN in \citet{Maiolino25}, together with the measured values for J0822$+$2241 that are fully consistent with the vast majority of measured upper limits for \textit{JWST}-detected type~1 AGN. We additionally include the broad H$\alpha$ luminosity versus 2\,--\,10\,keV luminosity relation of \citet{Shimizu18} for type~1\,--\,1.2 AGN detected in the 70-month BAT catalogue. The X-ray luminosity passband used in the relation was converted from 14\,--\,195\,keV to 2\,--\,10\,keV assuming a power law with a photon index of 1.8. To verify the X-ray passband conversion, we additionally show the narrow component subtracted H$\alpha$ luminosity versus 2\,--\,10\,keV luminosity relation found by \citet{Jin12b} for type~1 and Narrow Line Seyfert~1 AGN.

The overall disagreement of both the \textit{JWST} AGN and J0822$+$2241 with the relations of \citet{Shimizu18} and \citet{Jin12b} is clearly striking. J0822$+$2241 lies approximately $\sim$\,1\,--\,2\,dex below the relations and is consistent with the (predominantly) X-ray-undetected \textit{JWST} AGN. Given J0822$+$2241 is one of the brightest Green Pea galaxies observed in X-rays (e.g., \citealt{Svoboda19,Singha24}) though comparable in terms of broad H$\alpha$ luminosity \citep{Lin24a,Singha24}, AGN within other Green Pea galaxies would likely lie at even lower X-ray fluxes. Our X-ray analysis of J0822$+$2241 offers a possibility that Green Pea galaxies are X-ray analogues of the broad line \textit{JWST} AGN candidates. However, comparing a single source to the predominantly undetected X-ray sample of \citet{Maiolino25} is undoubtedly insufficient to draw conclusions on higher redshift objects. We additionally note that despite some similarities, the observed equivalent width of broad H$\alpha$ we detect for J0822$+$2241 is below the values found by \citet{Maiolino25} suggesting the possibility of high covering factor Compton-thick broad line region material in J0822$+$2241 to be particularly unlikely.

If future detailed X-ray spectral analyses of more Green Pea galaxies find analogous properties to J0822$+$2241 as well as the \textit{JWST}-detected broad line AGN, alternative scenarios may be required to explain their X-ray deficit. For example, a possible explanation for apparent flux suppression in the X-ray and/or optical bands is for the observed light to be scattered emission from dense obscuration surrounding the AGN. A scattered component is often invoked in X-ray spectral fitting of obscured AGN to explain an excess of flux at $\sim$\,2\,--\,5 keV (e.g., \citealt{Ricci17_bassV,Boorman25_nulands}). However, for a sufficiently Compton-thick AGN the scattered light component could dominate the entire X-ray spectral passband $<$\,10\,keV (e.g., \citealt{Greenwell22,Greenwell24}). Away from X-rays, \citet{Assef16,Assef20} studied a subset of Hot Dust Obscured Galaxies with unexpected blue ultraviolet-to-optical excesses, one of which was later confirmed to have a significant scattered component via imaging polarisation \citep{Assef22}. A significant scattered component could also potentially explain the broad H$\alpha$ emission detected in J0822$+$2241, although the intrinsic broad H$\alpha$ luminosity would have to be even higher than the current value which could exacerbate the effect seen in Figure~\ref{fig:gp2_jwst_agn}. The X-ray scattered fraction for Compton-thick AGN in the local universe is expected to be $\lesssim$\,1\% \citep{Gupta21} on average, such that the intrinsic X-ray luminosity of J0822$+$2241 would be $\gtrsim$\,10$^{44}$\,erg\,s$^{-1}$ if the observed \textit{XMM-Newton} spectra were dominated by a scattered component. This is consistent with the findings of \citet{Kawamuro19}, who previously showed that a Compton-thick AGN with an intrinsic luminosity of log\,$L_{2-10\,{\rm keV}}$\,/\,erg\,s$^{-1}$\,$\gtrsim$\,43.8 could self-consistently explain the near-to-mid infrared properties of J0822$+$2241, as well as the non-detection from a 20\,ks \textit{NuSTAR} observation. Future polarimetric and/or higher-sensitivity hard X-ray observations (e.g., with a mission concept like \textit{HEX-P}; \citealt{Madsen24}) of J0822$+$2241 as well as other Green Pea galaxies could robustly search for evidence of significant scattered components arising from heavily obscured AGN.

\section{Summary}\label{sec:summary}

We have presented the multi-epoch X-ray spectral analysis of SDSS\,J082247.66\,$+$224144.0. Our key findings are as follows:

\begin{enumerate}
    \item We find the observed 2\,--\,10\,keV luminosity from two epochs separated by $\sim$\,6.2\,years to be entirely consistent, with observed rest-frame values of log\,$L_{2-10\,{\rm keV}}$\,/\,erg\,s$^{-1}$\,=\,42.0\,$\pm$\,0.3 and 42.2\,$\pm$\,0.1 for epochs~1 and~2, respectively (c.f. ~Section~\ref{sec:results}). Such luminosities are seldom produced by sources other than an AGN, such that based on luminosity arguments alone we find that J0822$+$2241 to be a strong AGN candidate.

    \item We fit the archival SDSS optical spectrum of J0822$+$2241, finding a statistically significant requirement for a broad component to the H$\alpha$ line. We find a FWHM$_{{\rm H}\alpha,\,{\rm broad}}$\,=\,1360$^{+70}_{-100}$\,km\,s$^{-1}$. Via fitting of the H$\beta$ line and the corresponding Balmer decrement derived from the ratio of narrow H$\alpha$ to H$\beta$, we derive an absorption-corrected broad H$\alpha$ luminosity of log\,$L_{{\rm H}\alpha,\,{\rm broad,\,int}}$\,/\,erg\,s$^{-1}$\,=\,42.1$^{+0.3}_{-0.2}$. Assuming the pre-determined black hole mass relation of \citet{Reines15}, we estimate broad H$\alpha$-based black hole mass of M$_{{\rm BH,\,H}\alpha}$\,=\,8$_{-2}^{+3}$\,$\times$\,10$^{6}$\,M$_{\odot}$ (c.f.~Sections~\ref{subsec:optres} and~\ref{subsec:obscuredagn}).

    \item To complement our X-ray and optical spectral analyses, we additionally collate and fit the broadband SED of J0822$+$2241 from X-ray to mid-infrared wavelengths with the \texttt{Lightning} code. We find that an AGN component is required to dominate at X-ray wavelengths, with a major contribution also expected in the near-infrared regime. Our fits additionally find that the AGN may provide a significant contribution in the UV-to-optical regime, though further analysis would be needed to confirm. Finally, we derive black hole and stellar masses from \texttt{Lightning} that are fully consistent with the relation of \citet{Reines15} (c.f.~Sections~\ref{subsec:reslightning} and~\ref{sec:mstel}).
    
    \item We show that the soft band 0.5\,--\,2\,keV flux of J0822$+$2241 has decreased significantly between the two epochs. If arising from a line-of-sight column density variation around an intrinsically non-variable AGN, the column density would have increased by a factor of log($N_{\rm H,\,2}$\,/\,$N_{\rm H,\,1}$)\,=\,1.08$^{+0.66}_{-0.72}$. Since the inclusion of a variable obscurer leads to an intrinsic photon index that is fully consistent with the Seyfert population, and the general ubiquity of obscuration in AGN, we deem this scenario the most likely. Using a bolometric correction and the black hole mass constrained from our H$\alpha$ fitting, we estimate an Eddington ratio for J0822$+$2241 of $\lambda_{\rm Edd}$\,=\,1.4$_{-0.7}^{+0.4}$\,\%, potentially placing the source in (or close to) the unstable region of the effective Eddington limit on dusty gas (c.f.~Section~\ref{subsec:obscuredagn}).
    
    \item We find the possibility of the soft X-ray flux variability to have occurred from a single X-ray-unobscured AGN to be unlikely primarily based on the observed X-ray spectral shape changes between epoch~1 and~2. By considering a scenario in which the soft passband flux change were caused by a variable accretion disk component, we estimate the black hole mass would need to be $1.1_{-0.9}^{+36.0}$\,$\times$\,10$^{4}$\,M$_{\odot}$. However, given the super-Eddington accretion rates required to sustain the observed luminosity of J0822$+$2241, we deem our intermediate black hole mass estimate to be over-simplified and untrustworthy (c.f.~Section~\ref{subsec:unobscuredagn}).
    
    \item We investigate the possibility that the X-ray properties from J0822$+$2241 are the result of unresolved ULX and/or HLX sources in its host galaxy. If originating from a single off-nuclear source, J0822$+$2241 would be the highest-luminosity HLX ever discovered. We combine recently analysed radio flux measurements of J0822$+$2241 to derive a 5\,GHz-to-2\,--\,10\,keV luminosity ratio that is additionally fully-consistent with that expected from an AGN (c.f.~Section~\ref{subsec:ulxhlx}).

    \item We compare the broad H$\alpha$ and X-ray luminosities of J0822$+$2241 to the values measured for a sample of \textit{JWST}-detected type~1 AGN lacking detectable X-ray counterparts. We find the X-ray flux for J0822$+$2241 to be $\sim$\,1\,--\,2\,dex lower than expected using relations derived from local populations of AGN. The observed X-ray deficit and/or broad H$\alpha$ luminosity excess is in agreement with the (predominantly undetected) X-ray fluxes found for \textit{JWST}-detected broad line AGN. However, our X-ray spectral analysis of J0822$+$2241 does not find obvious evidence for Compton-thick obscuration at $<$\,10\,keV, nor intrinsically steep X-ray spectra that were proposed previously to explain the X-ray deficit observed in \textit{JWST}-detected AGN. We postulate that if future X-ray spectral analyses of Green Pea AGN and \textit{JWST} detected AGN show agreements, Green Pea galaxies may be useful for understanding black hole growth in compact galaxies in the early universe (c.f.~Section~\ref{sec:jwstcomp}).
    
\end{enumerate}

Out of all the scenarios tested in this work, an AGN displaying obscuration variability is the only appropriate possibility to explain the X-ray properties of J0822$+$2241 observed by \textit{XMM-Newton}. A fundamental requirement for this characterisation is that the AGN dominates the observed X-ray spectrum at $<$\,10\,keV. This is currently rare for the highly star forming galaxy population, with non-AGN related processes often concealing signatures of AGN activity $<$\,10\,keV (e.g., \citealt{Lehmer23,Brightman24}). Future high-sensitivity spectroscopy above 10\,keV with capabilities like the \textit{High Energy X-ray Probe} (\textit{HEX-P}; \citealt{Madsen24}) would be a powerful tool for correctly characterising the census of AGN within low mass low metallicity galaxies (including more extreme luminous compact galaxies) unveiled en masse by next-generation missions such as the \textit{UltraViolet Explorer} (\textit{UVEX}; \citealt{Kulkarni21}). More broadly, given the current obstacles associated with detecting high-redshift \textit{JWST}-detected AGN candidates in X-rays (e.g., \citealt{Ananna24,Yue24,Maiolino25}), our work shows that future dedicated X-ray campaigns of local analogues may prove fruitful for a complete understanding of black hole growth at high redshift.



\section*{ACKNOWLEDGEMENTS}
We thank the anonymous reviewer for their useful comments that helped to improve the manuscript. P.G.B., J.S., A.B. \& K.K. acknowledge financial support from the Czech Science Foundation under Project No. 22-22643S. B.A. acknowledges financial support from Charles University under GAUK Project No. 345625. The work of D.S. was carried out at the Jet Propulsion Laboratory, California Institute of Technology, under a contract with NASA. R.J.A. was supported by FONDECYT grant number 1231718 and by the ANID BASAL project FB210003. D.J.W. acknowledges support from the Science and Technology Facilities Council (STFC; grant code ST/Y001060/1).

This research has made use of the NASA/IPAC Extragalactic Database (NED), which is operated by the Jet Propulsion Laboratory, California Institute of Technology, under contract with the National Aeronautics and Space Administration.

This research has made use of NASA’s Astrophysics Data System Bibliographic Services.


%

\vspace{5mm}
\facilities{\textit{XMM-Newton}}


\software{This paper made extensive use of \texttt{matplotlib} \citep{Hunter2007}, \texttt{pandas} \citep{reback2020pandas, mckinney-proc-scipy-2010} and \texttt{astropy} \citep{astropy:2013,astropy:2018,astropy:2022}.}



\appendix
\section{Updated Palomar spectroscopy}\label{app1:palomar}
\begin{figure}
\centering
\includegraphics[width=0.99\textwidth]{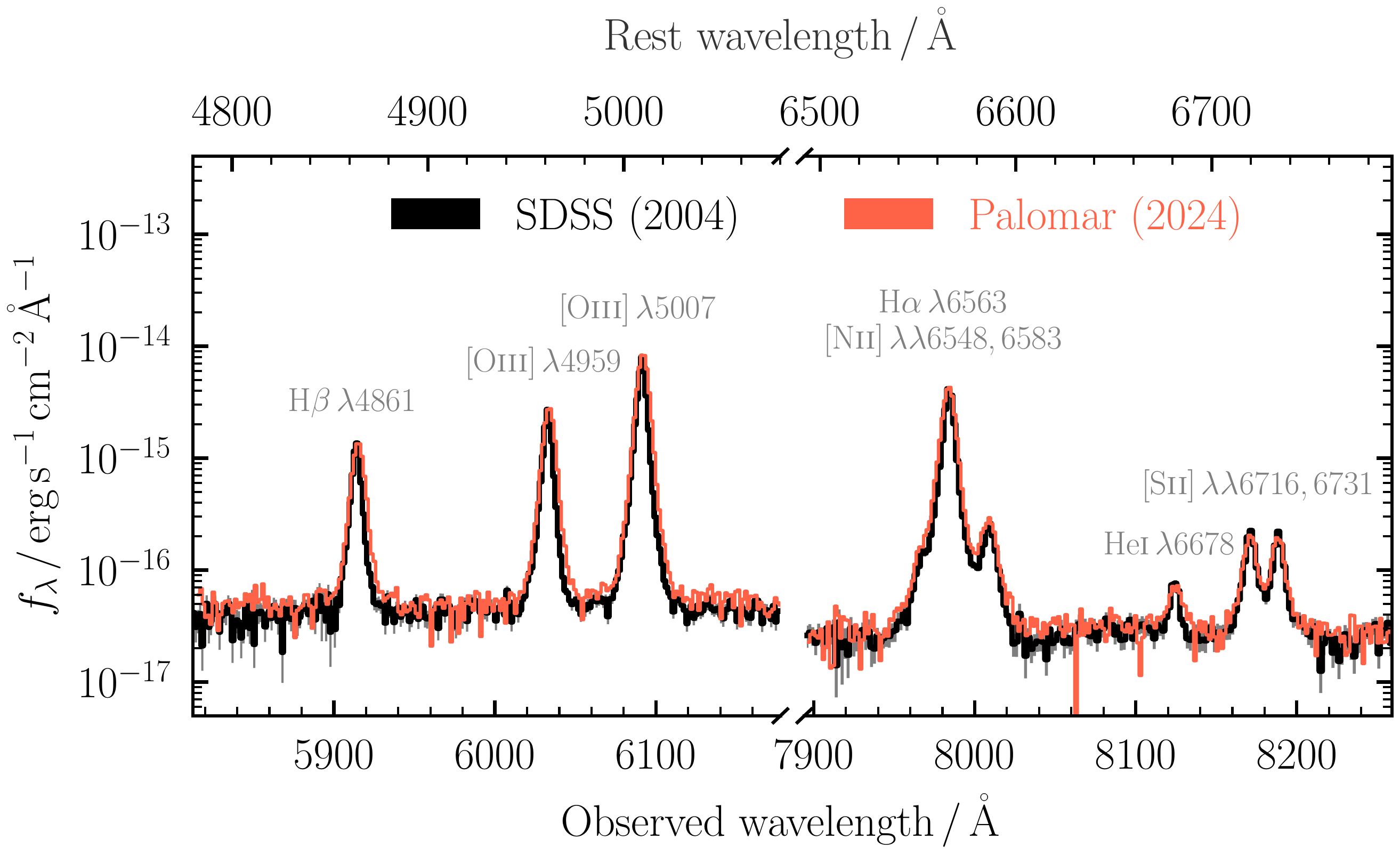}
\caption{Comparison between the line profiles measured by SDSS and Palomar/DoubleSpec with thick black and thinner red lines, respectively. From left to right, we show H$\beta$\,$\lambda$4861, [O\textsc{ii}]\,$\lambda\lambda$4959,\,5507, H$\alpha$\,$\lambda$6563, [N\textsc{ii}]\,$\lambda\lambda$6548,\,6583, He\textsc{i}\,$\lambda$6678 and [S\textsc{ii}]\,$\lambda\lambda$6716,\,6731. After applying a simple translation in wavelength and flux, the observed spectra are considerably similar , shifted by comparable amounts in relative flux and absolute wavelength. Even without applying a translation to the Palomar spectrum, all line profiles are qualitatively similar, suggesting a lack of spectral variability for the H$\alpha$ line profile over an $\sim$\,16 year rest-frame baseline.}\label{fig:gp2_palomar_comp}
\end{figure}

As a means to investigate the temporal properties of the optical spectrum of J0822$+$2241, we acquired follow-up spectroscopy with the Palomar/DoubleSpec spectrograph on UT\,2024\,Oct\,04. Given the SDSS spectrum was taken in 2004, our follow-up Palomar spectroscopy covers an observed baseline of $\sim$\,20\,years or $\sim$\,16\,years in the rest-frame of J0822$+$2241. We apply an additional linear correction to wavelength and flux of the Palomar spectrum in order to match the observed narrow lines in the spectrum. We find an acceptable match in the [O\textsc{ii}]\,$\lambda\lambda$4959,\,5507 and [S\textsc{ii}]\,$\lambda\lambda$6716,\,6731 narrow lines by shifting the spectrum 2.1\,${\rm \AA}$ redward and increasing the observed flux by a multiplicative factor of 2.1. The original SDSS and correct Palomar spectra are over-plotted in Figure~\ref{fig:gp2_palomar_comp}. There are no obvious changes in any key line profiles, in particular for the H$\alpha$\,$\lambda$6563\,$+$\,[N\textsc{ii}]\,$\lambda\lambda$6548,\,6583. The consistency over an $\sim$\,16\,year baseline strongly suggests the broad component to H$\alpha$ that we detect cannot arise from a single supernova, for which the broad component would be expected to decay on the order of years (e.g., \citealt{Izotov07}).

\bibliography{bibliography}{}
\bibliographystyle{aasjournal}



\end{document}